\documentclass[notitlepage,superscriptaddress,nofootinbib,tightenlines,twocolumn]{revtex4-1}
\usepackage{amsmath,verbatim,latexsym,amssymb,indentfirst,mathrsfs,mathtools,amsthm,bbm,bm,hyperref,url,cancel,subcaption,enumitem}
\usepackage[font=small,labelfont=bf,format=plain,justification=raggedright,singlelinecheck=false]{caption}
\usepackage{verbatim,indentfirst}
\usepackage[title,titletoc]{appendix}

\usepackage{graphicx}
\usepackage{epstopdf}
\newcommand{\caphead}[1]{{\bf #1}}

\renewcommand{\thesection}{\Roman{section}}
\renewcommand{\thesubsection}{\Roman{section} \Alph{subsection}}
\renewcommand{\thesubsubsection}{\Roman{section} \Alph{subsection} \arabic{subsubsection}}
\makeatletter
\def\p@subsection{}
\makeatother
\makeatletter
\def\p@subsubsection{}
\makeatother

\makeatletter
\newcommand\footnoteref[1]{\protected@xdef\@thefnmark{\ref{#1}}\@footnotemark}
\makeatother

\usepackage{soul}

\usepackage{color}
\usepackage[dvipsnames]{xcolor}
\usepackage[normalem]{ulem}

\newcommand{\Sites}{N}  
\newcommand{\sites}{n}  

\newcommand{\amc}{\mathcal{M}}
\newcommand{\triv}{{\rm triv}}

\newcommand{\NATS}{{\rm NATS}}
\newcommand{\can}{{\rm can}}
\newcommand{\GC}{{\rm GC}}

\newcommand{\numTrials}{\mathcal{N}_{\rm trials}}  
\newcommand{\BH}{{\rm BH}}

\newcommand{\inter}{ {\rm int} }   
\newcommand{\hc}{ {\rm h.c.} }

\newcommand{\tot}{ {\rm tot} }
\def\const{ {\rm const.} }   
\newcommand{\Tr}{{\rm Tr}}   
\def\id{\mathbbm{1}}   

\DeclareMathOperator{\supp}{supp}

\newcommand{\Sys}{{\rm S}}  
\newcommand{\Bath}{{\rm B}} 

\newcommand{\JParen}{ {(j)} }

\newcommand{\LParen}{ \bm{(} }
\newcommand{\RParen}{ \bm{)} }

\newcommand*{\Set}[1]{\left\{  #1  \right\}}

\renewcommand\th{ {\rm th} }

\newcommand*{\ket}[1]{\lvert #1 \rangle}

\newcommand*{\ketbra}[2]{\lvert #1 \rangle\!\langle #2 \rvert}
\newcommand*{\expval}[1]{\left\langle  #1  \right\rangle}

\begin{document}
\title{
Noncommuting conserved quantities in quantum many-body thermalization}
%
\author{Nicole~Yunger~Halpern}
\email{nicoleyh@g.harvard.edu}
\affiliation{Institute for Quantum Information and Matter, California Institute of Technology, Pasadena, CA 91125, USA}
\affiliation{ITAMP, Harvard-Smithsonian Center for Astrophysics, Cambridge, MA 02138, USA}
\affiliation{Department of Physics, Harvard University, Cambridge, MA 02138, USA}
\affiliation{Research Laboratory of Electronics, Massachusetts Institute of Technology, Cambridge, Massachusetts 02139, USA}
\author{Michael~E.~Beverland}
\affiliation{Microsoft Quantum, Redmond, WA, USA}
\author{Amir Kalev}
\email{amirk@umd.edu}
\affiliation{Joint Center for Quantum Information and Computer Science, University of Maryland, College Park, MD 20742-2420, USA}
\date{\today}

%
%
\begin{abstract}
In statistical mechanics, a small system exchanges conserved quantities---heat, particles, electric charge, etc.---with a bath. The small system thermalizes to the canonical ensemble, or the grand canonical ensemble, etc., depending on the conserved quantities. The conserved quantities are represented by operators usually assumed to commute with each other. This assumption was removed within quantum-information-theoretic (QI-theoretic) thermodynamics recently. The small system's long-time state was dubbed ``the non-Abelian thermal state (NATS).'' We propose an experimental protocol for observing a system thermalize to the NATS. We illustrate with a chain of spins, a subset of which form the system of interest. The conserved quantities manifest as spin components. Heisenberg interactions push the conserved quantities between the system and the effective bath, the rest of the chain. We predict long-time expectation values, extending the NATS theory from abstract idealization to finite systems that thermalize with finite couplings for finite times. Numerical simulations support the analytics: The system thermalizes to the NATS, rather than to the canonical prediction. Our proposal can be implemented with ultracold atoms, nitrogen-vacancy centers, trapped ions, quantum dots, and perhaps nuclear magnetic resonance. This work introduces noncommuting conserved quantities from QI-theoretic thermodynamics into quantum many-body physics: atomic, molecular, and optical physics and condensed matter.
\end{abstract}

{\let\newpage\relax\maketitle}

Quantum noncommutation was recently introduced into
the following textbook statistical mechanics problem:
Consider a small quantum system exchanging heat with 
a large bath via weak coupling.
The small system equilibrates to 
a canonical ensemble~\cite{Laundau_80_Statistical},
\begin{align}
   \label{eq_Rho_Can}
   \rho_\can
   :=  e^{ - \beta H^\Sys }  /  Z_\can^\Sys .
\end{align}
$\beta = 1 / T$ denotes the bath's inverse temperature
(we set Boltzmann's constant to one),
$H^\Sys$ denotes the system-of-interest Hamiltonian, and
the partition function $Z_\can^\Sys  :=  \Tr  (  e^{ - \beta H^\Sys }  )$
normalizes the state.
If the system and bath exchange heat and particles,
the system equilibrates to a grand canonical ensemble
$\propto  e^{ - \beta (H^\Sys  -  \mu N^\Sys) }$.
The bath's chemical potential is denoted by $\mu$,
and $N^\Sys$ denotes the system-of-interest particle-number operator.
This pattern extends to electric charge 
and other globally conserved extensive quantities.
We call the quantities \emph{charges},
even when referring to the nonconserved system or bath charges,
for convenience.
The charges are represented by Hermitian operators
assumed implicitly to commute with each other.

A few references addressed this assumption during the 20th century:
Jaynes and followers applied the principle of maximum of entropy
to charges that fail to commute with each other~\cite{Jaynes_57_Information_II,Balian_87_Equiprobability,Balian_86_Dissipation}.
These initial steps drive us to ask what fails to hold---how 
charges' noncommutation alters thermalization and transport---and 
to complement Jaynes's information-theoretic treatment 
with physical treatments of thermalization.

First steps were taken 
within quantum-information-theoretic (QI-theoretic) thermodynamics~\cite{NYH_18_Beyond,Lostaglio_14_Masters,Lostaglio_14_Masters,NYH_16_Microcanonical,Guryanova_16_Thermodynamics,Lostaglio_17_Thermodynamic}.
A small system was imagined to exchange with a bath
charges $Q_\alpha$ that need not commute:
$[ Q_\alpha ,  Q_{\alpha'} ]  \neq  0$.
Whether the system of interest can thermalize is unclear, 
for three reasons.
First, a small system thermalizes
if the global system is prepared in a microcanonical subspace,
an eigenspace shared by the global charges.
If the global charges fail to commute,
they do not necessarily share a degenerate eigenspace,
so a microcanonical subspace might not exist.
Second, the total Hamiltonian $H^\tot$ 
conserves each total charge $Q_\alpha^\tot$,
so $H^\tot$ shares an eigenbasis with each $Q_\alpha^\tot$.
But the $Q_\alpha^\tot$'s do not commute,
so they do not share an eigenbasis.
Hence $H^\tot$ may have an unusual degeneracy pattern,
and degeneracies tend to nullify expectations about thermalization.
Third, noncommutation invalidates a derivation of 
the thermal state's form~\cite{NYH_16_Microcanonical}.
Appendix~\ref{sec_U1} details further how charges' noncommutation
invalidates the eigenstate thermalization hypothesis (ETH),
which elucidates why chaotic quantum many-body systems thermalize internally~\cite{Deutsch_91_Quantum,Srednicki_94_Chaos,Rigol_08_Thermalization,D'Alessio_16_From}.
Similarly, Sec.~\ref{sec_Discussion} distinguishes 
the NATS from the generalized Gibbs ensemble (GGE),
to which integrable systems equilibrate~\cite{Rigol_09_Breakdown,Rigol_07_Relaxation,Vidmar_16_Generalized}.

QI theory was deployed to argue that 
a thermal state exists and has the form~\cite{NYH_16_Microcanonical,Guryanova_16_Thermodynamics,Lostaglio_17_Thermodynamic} 
\begin{align}
   \label{eq_NATS}
   \rho_\NATS  :=
   e^{ -  \beta  \left( H^\Sys  
                              -  \sum_{\alpha = 1}^c  \mu_\alpha  Q_\alpha^\Sys  \right)  
        }  / Z_\NATS^\Sys .
\end{align}
$H^\Sys$ denotes the system-of-interest Hamiltonian,
$Q_\alpha^\Sys$ denotes the $\alpha^\th$
system-of-interest charge,
the $\mu_\alpha$'s denote generalized chemical potentials,
and the partition function 
$Z_\NATS^\Sys  :=  \Tr  \left(  
e^{ -  \beta  \left( H^\Sys  
     -  \sum_\alpha  \mu_\alpha  Q_\alpha^\Sys  \right)  }  \right)$
normalizes the state.
Though Eq.~\eqref{eq_NATS} has the expected exponential form,
fully justifying this form requires considerable mathematical effort
when the charges fail to commute~\cite{NYH_18_Beyond,Lostaglio_14_Masters,Lostaglio_14_Masters,NYH_16_Microcanonical,Guryanova_16_Thermodynamics,Lostaglio_17_Thermodynamic}.
This \emph{non-Abelian thermal state} (NATS)~\eqref{eq_NATS}~\cite{NYH_16_Microcanonical}
has since spread across QI-theoretic thermodynamics~\cite{Ito_18_Optimal,Bera_17_Thermodynamics,Mur_Petit_18_Revealing,Gour_18_Quantum,Popescu_18_Quantum,Manzano_18_Squeezed,Sparaciari_18_First}.

This QI-theoretic approach offers the benefits
of mathematical precision and cleanliness.
Yet this abstract, formal, idealized approach
is divorced from implementations.
Whether any real physical system could exchange noncommuting charges,
what the system could consist of,
how the charges would manifest,
which interactions could implement the exchange, etc.
have been unknown.
Can NATS physics exist outside of mathematical physics?

We answer this question affirmatively,
arguing that the NATS theory of QI-theoretic thermodynamics
follows from infinitely long thermalization
at infinitely weak coupling.
We show how to realize the NATS under realistic conditions
in condensed-matter, atomic-molecular-and-optical (AMO), 
and high-energy systems.
These fields have recently experienced a surge of interest
in many-body thermalization.
Therefore, we propose and numerically simulate an experimental protocol 
for observing a quantum many-body system thermalize to the NATS,
a peculiarly nonclassical thermal state
that has never been observed.
Our protocol is suited to cold and ultracold atoms~\cite{Bakr_09_Quantum,Sherson_10_Single,Parsons_15_Site,Omran_15_Microscopic,Cheuk_15_Quantum,Barredo_16_Atom,Endres_16_Atom,Gring_12_Relaxation,Bernien_17_Probing,Prufer_18_Observation,Kaufman_16_Quantum,deLeseleuc_18_Experimental}
%
superconducting qubits~\cite{Neill_16_Ergodic,Devoret_13_Superconducting,Majer_07_Coupling}, 
trapped ions~\cite{Smith_16_Many,Zhang_17_Observation,Brydges_19_Probing},
%
nitrogen-vacancy centers in diamond~\cite{Kucsko_18_Critical},
quantum dots~\cite{Kandel_19_Coherent,Hensgens_17_Quantum},
and perhaps nuclear magnetic resonance (NMR)~\cite{Wei_18_Emergent}.
To extend the NATS theory to finite times and coupling strengths,
we propose initial steps toward a NATS many-body theory.
The point is that enhancing undergraduate statistical mechanics---the 
grand canonical ensemble---with noncommuting charges
produces a thermal state that has never been observed.
We observe such thermalization numerically,
and we propose an experimental observation.
The proposal shares the spirit of
observations of the GGE~\cite{Langen_15_Experimental}.

\begin{figure}[hbt]
\centering
\includegraphics[width=.35\textwidth, clip=true]{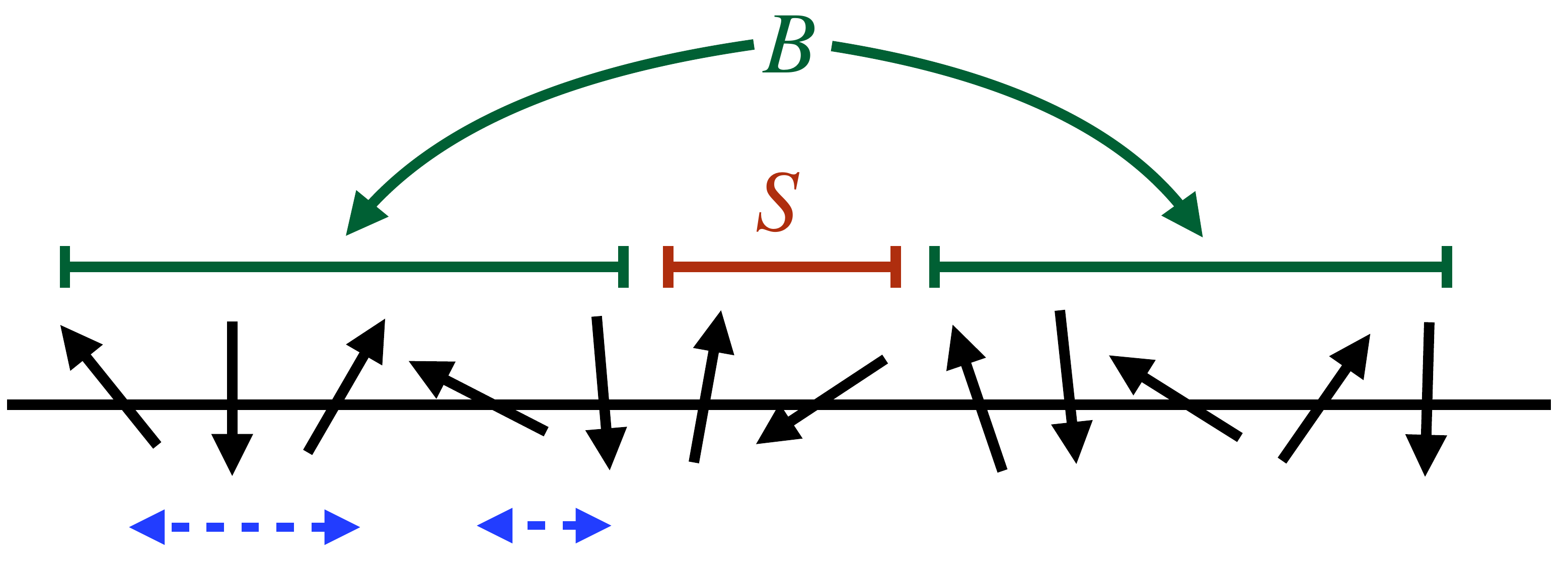}
\caption{\caphead{Setup for thermalization to 
the non-Abelian thermal state (NATS):} 
We illustrate the general experimental setup with 
the spin-chain example proposed in Sec.~\ref{sec_Proposal}.
The system $\Sys$ of interest consists of $\sites = 2$ qubits.
The other qubits form an effective bath $\Bath$.
The dashed, blue arrows illustrate nearest-neighbor
and next-nearest-neighbor interactions.}
\label{fig_Setup}
\end{figure}

The many-body NATS theory can be tested with 
an experiment of the following form.
Consider a closed, isolated set of 
$\Sites$ identical copies of a quantum system.
We illustrate with a chain of qubits (quantum two-level systems),
realizable with ultracold atoms (Fig.~\ref{fig_Setup}).
One copy forms the system $\Sys$ of interest
(e.g., $\sites = 2$ qubits).
The other copies form an effective bath $\Bath$
(e.g., $\Sites - 1$ qubit pairs).

Copy $j$ evolves under a Hamiltonian $H^\JParen  =  H^\Sys$
and has charges $Q_\alpha^\JParen$
that fail to commute with each other.
In the spin-chain example, each 
$Q_\alpha^\JParen = \sigma_{\alpha = x, y, z}^\JParen$
manifests as a spin component.
We neglect factors of $1/2$ and set $\hbar = 1$.
$H^\JParen$ preserves each local charge,
in analogy with the grand canonical problem:
There, if the system is isolated from the bath,
the system's particle number remains constant.
An interaction Hamiltonian $H^\inter$
pushes charges between $\Sys$ and $\Bath$.
$H^\inter$ conserves each total charge,
$Q_\alpha^\tot  
:=  \sum_{j = 1}^\Sites  Q_\alpha^\JParen$.
The total Hamiltonian, 
$H^\tot  :=  \sum_{j = 1}^\Sites  H^\JParen  +  H^\inter$,
is nonintegrable, to promote thermalization.
The total Hamiltonian preserves each total charge:
$[H^\tot,  Q_\alpha^\tot]  =  0$.
We illustrate $H^\tot$ with nearest-neighbor 
and next-nearest-neighbor Heisenberg interactions.

%

The whole system is prepared in a state $\rho$
in which each total charge has a fairly well-defined value:
Measuring any $Q_\alpha^\tot$ or $H^\tot$ has a high probability
of yielding a value close to the ``expected value,'' 
$S_\alpha$ or $E^\tot$.
$S_\alpha$ and $E^\tot$ serve analogously to 
the grand canonical problem's $N^\tot$ and $E^\tot$.
The whole system then thermalizes internally
for a long time under $H^\tot$.
A time linear in the system size suffices,
according to numerics.
A local observable $O$ of $\Sys$ is then measured.

We posit that the expectation value thermalizes to
\begin{align}
   \label{eq_O_Predict_Intro}
   \Tr \left(  O  
   e^{ - \beta \left( H^\tot  -  \sum_\alpha  \mu_\alpha  Q_\alpha^\tot  \right) }
   /  Z_\NATS^\tot  \right) .
\end{align}
$\beta$ and the $\mu_\alpha$'s, we posit, 
depend on $E^\tot$ and the $S_\alpha$'s through
\begin{align}
   \label{eq_Beta_Predict_Intro}
   & E^\tot  
   =  \Tr \left(  H^\tot
                      e^{ - \beta \left( H^\tot - \sum_\alpha \mu_\alpha Q_\alpha^\tot  \right) 
                          }  \right)  /  Z_\NATS^\tot
   \; \; \text{and} \\
   \label{eq_Mu_Predict_Intro}
   & S_\alpha  
   =  \Tr \left(  Q_\alpha^\tot
                      e^{ - \beta \left( H^\tot - \sum_\alpha \mu_\alpha Q_\alpha^\tot  \right) 
                          }  \right)  /  Z_\NATS^\tot  .
\end{align}
These equations parallel the definition of inverse temperature, $\beta$, 
in many-body studies of energy conservation~\cite{D'Alessio_16_From}.
We calculate $\beta$ and the $\mu$'s analytically 
in the spin-chain example.

En route to the thermodynamic limit, 
$\Sys$ and $\Bath$ grow large.
The characteristic scale of $H^\inter$ remains constant,
while the scale of $H^\Sys$ grows.
The scales' ratio approaches zero.
The whole-system quantities in Eq.~\eqref{eq_O_Predict_Intro} 
can be replaced with $\Sys$ quantities:
\begin{align}
   & \Tr \left(  O  
   e^{ - \beta \left( H^\tot  -  \sum_\alpha  \mu_\alpha  Q_\alpha^\tot  \right) }
   /  Z_\NATS^\tot  \right)
   \nonumber \\ & \quad 
   \label{eq_O_Predict_Intro_2}
   \to  \Tr  \left(  O
   e^{ - \beta  \left( H^\Sys  -  \sum_\alpha  \mu_\alpha  Q_\alpha^\Sys
         \right)  }  /  Z_\NATS^\Sys  \right) .
\end{align}
Let $\rho_\Sys$ denote the long-time state (reduced density operator) of $\Sys$.
If all $\Sys$ observables $O$ 
thermalize as in~\eqref{eq_O_Predict_Intro_2},
$\Sys$ thermalizes to the NATS~\eqref{eq_NATS}
of idealized QI-theoretic thermodynamics.
Numerical simulations confirm that 
the state approaches the NATS prediction:
\begin{align}
   \rho_\Sys
   \approx  \rho_\NATS .
\end{align}

The rest of this paper is organized as follows.
Section~\ref{sec_Proposal} illustrates our experimental proposal
with a spin chain realizable with, e.g., ultracold atoms.
Numerical simulations in Sec.~\ref{sec_Numerics}
support the analytical predictions.
Section~\ref{sec_Discussion} presents opportunities
created by the introduction of noncommuting charges 
into many-body thermalization.

\section{Proposal for spin-chain experiment}
\label{sec_Proposal}

We sketched a general experimental protocol in the introduction.
Here, we illustrate with a spin chain.
We detail the setup (Sec.~\ref{sec_Setup}), 
preparation procedure (Sec.~\ref{sec_Prep}), 
evolution (Sec.~\ref{sec_Evolve}), 
and readout (Sec.~\ref{sec_Readout}).

\subsection{Setup}
\label{sec_Setup}

Let $\Sys$ denote a system of $\sites > 1$ qubits.
Consider a chain of $\Sites$ copies of $\Sys$
(Fig.~\ref{fig_Setup}).
A multidimensional lattice would suffice,
as discussed before Eq.~\eqref{eq_HTot}.
The non-$\Sys$ copies form the effective bath, $\Bath$.
We index the qubits with $j = 1, 2, \ldots, \Sites \sites$ 
and the subsystems with $k = 1, 2, \ldots, \Sites$.

Let $\sigma_\alpha^\JParen$ denote
component $\alpha = x, y, z$ of the qubit-$j$ spin.
The spin operators satisfy the eigenvalue equations
$\sigma_\alpha^\JParen  
\ket{ \alpha \pm }_j
=  \pm  \ket{ \alpha \pm }_j$.
The chain has the total spin
$\sigma_\alpha^\tot
:=  \sum_{j = 1}^{\Sites \sites}
\sigma_\alpha^\JParen$.
Spin was applied in quantum thermodynamics 
to work extraction previously~\cite{Vaccaro_11_Information,Wright_18_Quantum,NYH_18_Beyond,Guryanova_16_Thermodynamics,NYH_16_Microcanonical,Lostaglio_17_Thermodynamic}.

$H^\tot$ must conserve each $\sigma_\alpha^\tot$
while transferring subsystem charges between $\Sys$ and $\Bath$.
We construct such an $H^\tot$ through physical reasoning.
Let $\sigma_{\pm \alpha}^\JParen$ denote 
the raising and lowering operators
for component $\alpha$ of the qubit-$j$ spin.
For example, $\sigma_{\pm z}^\JParen
= \frac{1}{2} (  
\sigma_x^\JParen  \pm  i \sigma_y^\JParen  )$.
Rotating each side of this equation unitarily yields
the raising and lowering operators for
components $x$ and $y$.
The two-site operator
$J_\alpha  ( \sigma_{+ \alpha}^\JParen  \sigma_{- \alpha}^{(j+1)}
+ \hc )$ transports $\alpha$-charges between sites $j$ and $j+1$
with frequency $J_\alpha$.
The Hamiltonian must transport charges of all types,
so we sum over $\alpha$.
For the Hamiltonian to commute with each total charge,
the $J_\alpha$'s must equal each other.
A Heisenberg interaction results:
$J \sum_{\alpha = x, y, z}  
   (\sigma_{+ \alpha}^\JParen  \sigma_{- \alpha}^{(j+1)}
     + \hc  )
   =  J  \vec{\sigma}^\JParen  \cdot  \vec{\sigma}^{(j+1)}$.

The nearest-neighbor Heisenberg interaction is integrable.
Next-nearest-neighbor interactions break integrability,
as would a higher-dimensional lattice.
We therefore choose for the spin chain to evolve under
\begin{align}
   \label{eq_HTot}
   H^\tot
   & =  J  \Bigg(
   \sum_{j = 1}^{\Sites \sites - 1} 
   \vec{\sigma}^\JParen  \cdot  \vec{\sigma}^{(j+1)} 
   +  \sum_{j = 1}^{\Sites \sites - 2} 
   \vec{\sigma}^\JParen  \cdot  \vec{\sigma}^{(j+2)} 
   \Bigg)  .
\end{align}
Similar interactions have been realized with
ultracold atoms~\cite{Fukuhara_13_Microscopic,Barredo_16_Atom,deLeseleuc_18_Experimental},
%
symmetric top molecules~\cite{Wall_15_Realizing,Glaetzle_15_Designing},
trapped ions~\cite{Zhang_17_Observation},
and NMR~\cite{Wei_18_Emergent}.
Furthermore, anisotropic interactions can be used to
generate isotropic effective interactions~\cite{Viola_99_Universal,Jane_03_Simulation}, as follows.
The system is evolved under the original Hamiltonian,
the $z$-axis is rotated into the $x$-axis,
the system is evolved further,
the new $x$-axis is rotated into the new $y$-axis,
and then the system is evolved further.

Our interaction is weak:
$H^\inter$ consists of the six $\vec{\sigma} \cdot \vec{\sigma}$ 
terms that link $\Sys$ to $\Bath$.
Hence the interaction energy $\sim 6J$.
$H^\Sys$ consists of the six bonds that act on just $\Sys$.
Hence $\Sys$ has energy $\sim J \sites$.
The interaction-energy-to-$\Sys$-energy ratio 
vanishes in the thermodynamic limit, as
$\Sites, \sites \to \infty$.
Hence the interaction is weak.
It is also when master equations predict equilibration 
to thermal states in the absence of noncommuting charges~\cite{Breuer_03_Theory,Cuetara_16_Quantum},
Hence one should expect $\Sys$ to thermalize.

\subsection{Preparation procedure}
\label{sec_Prep}
 
The grand canonical problem motivates our preparation procedure.
Consider aiming to watch a small system thermalize to
the grand canonical ensemble.
The system-and-bath composite should be prepared with
a well-defined total energy, $E^\tot$, 
and total particle number, $N^\tot$.
If classical, the whole system occupies a shell in phase space.
The shell's width stems from measurement imprecision.
If quantum, the total system approximately occupies 
a microcanonical subspace,
an eigenspace shared by the total Hamiltonian
and total particle-number operator~\cite{Laundau_80_Statistical,Popescu_06_Entanglement}.

Let us translate this protocol into the noncommuting problem.
One might aim to prepare the whole system with
a well-defined $\sigma_\alpha^\tot$ for all $\alpha = x, y, z$.
But the spin components fail to commute;
they share no joint eigenspace.
The microcanonical subspace was therefore generalized to an 
\emph{approximate microcanonical (a.m.c.) subspace,}
$\amc$~\cite{NYH_16_Microcanonical}.
In $\amc$, every total charge has 
a fairly well-defined value $S_\alpha$.
We propose two protocols for preparing the global system
in a state that occupies an a.m.c. subspace.
Measuring any $\sigma_\alpha^\tot$ will have 
a high probability of yielding a value close to $S_\alpha$.
$S_\alpha$ serves similarly to 
the commuting problem's $N^\tot$.
The probability and closeness were quantified in~\cite{NYH_16_Microcanonical}
and are reviewed below.
The longer the spin chain,
the more certain the measurement outcome can be.

We seek to prepare an initial global state that exhibits at least two properties:
\begin{enumerate}[label=(\roman*)]

   \item \label{item_Std_Dev}
   Each total charge has a standard deviation bounded as
\begin{align}
   \label{eq_Std_Devn_Condn} 
   \sqrt{  \expval{ ( \sigma_\alpha^\tot )^2 }
   -  \expval{ \sigma_\alpha^\tot }^2  }
   \sim O \left( [ \Sites  \sites ]^a \right) ,
   \; \text{wherein} \;
   a \leq 1/2 .
\end{align}

    \item \label{item_Not_Eigenstate}
    The initial global state is not an $H^\tot$ eigenstate.
For example, the spins do not all point in the same direction.
\end{enumerate}
We exhibit two protocols that satisfy conditions~\ref{item_Std_Dev}
and~\ref{item_Not_Eigenstate},
a \emph{product-state protocol} and
a \emph{soft-measurement protocol}.
Additionally, certain spin squeezed states~\cite{Kitagawa_93_Squeezed,Wineland_94_Squeezed} 
may be able to serve as initial states.

\emph{Product-state protocol:} 
A fraction $S_\alpha / (\Sites \sites)$ of the qubits 
are prepared in $\ket{ \alpha+ }$, 
for each of $\alpha = x, y, z$.
The state exhibits property~\ref{item_Std_Dev} because
every $\sigma_\alpha^\tot$ has
a subextensive standard deviation
in every short-range-correlated state
(App.~\ref{sec_Variance}).
%
The state's satisfaction of condition~\ref{item_Not_Eigenstate} 
was checked numerically.

\emph{Soft-measurement protocol:}
For motivation, we return to the grand canonical problem.
One can fix $E^\tot$ and $N^\tot$ by measuring 
the total energy, then the total particle number.
One could analogously, in the noncommuting problem, measure  
$H^\tot$, then $\sigma_x^\tot$, then $\sigma_y^\tot$, 
then $\sigma_z^\tot$.
But the $y$ and $z$ measurements would disturb
the $x$ and $y$ components.
The projective measurements must be ``softened.''
We define a \emph{soft measurement} as having two properties,
(a) \emph{peaking} and (b) \emph{mild disturbance}:
(a) Suppose that $\sigma_\alpha^\tot$ is measured softly,
yielding outcome $\tilde{S}_\alpha$.
Suppose that $\sigma_\alpha^\tot$ is then measured strongly.
The outcome must have a high probability 
of lying close to $\tilde{S}_\alpha$.
(b) Suppose that $\sigma_\alpha^\tot$ is measured strongly,
then some other $\sigma_{\alpha'}^\tot$ is measured softly,
and then $\sigma_\alpha^\tot$ is measured strongly again.
The final measurement must have a high probability
of yielding the first measurement's outcome.
The soft measurement must scarcely disturb 
$\sigma_\alpha^\tot$.

We formalize soft measurements in App.~\ref{sec_Soft_Meas},
using a positive operator-valued measure 
(a mathematical model for a generalized measurement~\cite{NielsenC10})
with a binomial envelope.
Similar measurements have been implemented
via weak coupling of system and detector~\cite[Eq.~(22)]{Jacobs_06_Straightforward}.
The soft measurements' ``peaking'' property 
determines the protocol's $S_\alpha$'s.
Mild disturbance ensures that the global charges
have small standard deviations, or property~\ref{item_Std_Dev}.
This property is checked numerically, at infinite temperature,
in App.~\ref{app_Soft_Meas_Std_Devn}.
Condition~\ref{item_Not_Eigenstate} was checked numerically.
Appendix~\ref{sec_Params} reconciles Ineq.~\eqref{eq_Std_Devn_Condn}
with the a.m.c. subspace's original definition~\cite{NYH_16_Microcanonical}.


%
%
%
\subsection{Evolution}
\label{sec_Evolve}

The whole system has been prepared in some state
in an a.m.c. subspace $\amc$.
The chain is now evolved under $H^\tot$.
Numerical simulations imply that
a time $\sim \Sites \sites / J$ suffices for distinguishing 
the NATS from the canonical prediction.
The interaction hops spin quanta between sites.
The evolution is intended to prepare the chain in
an \emph{a.m.c. ensemble},
the noncommuting analog of the microcanonical ensemble:
Let $P_\amc$ denote the projector onto $\amc$.
The a.m.c. ensemble is defined as 
$P_\amc / \Tr ( P_\amc )$~\cite{NYH_16_Microcanonical}.
Tracing out the bath from $P_\amc / \Tr ( P_\amc )$ 
was proved analytically to yield a system-of-interest state
close to the NATS~\cite{NYH_16_Microcanonical}.


%
%
%
\subsection{Readout}
\label{sec_Readout}
 
We aim to test experimentally the analytical prediction in~\cite{NYH_16_Microcanonical}.
Let $\rho_\Sys$ denote the long-time state of $\Sys$.
We posit that most local observables $O$ 
end with expectation values 
given by the NATS prediction~\eqref{eq_O_Predict_Intro}.
Equations~\eqref{eq_Beta_Predict_Intro} and~\eqref{eq_Mu_Predict_Intro} 
determine $\beta$ and the $\mu_\alpha$'s.

If the system is hot and the effective chemical potentials are small,
$\beta$ and the $\mu_\alpha$'s
can be calculated perturbatively (App.~\ref{sec_Calc_Mus}).
Loosely speaking, the assumptions are
\begin{align}
   \label{eq_High_T} &
   \sqrt{\Sites \sites}  \:  | \beta |  J ,  \; \;
   \sqrt{ \Sites \sites  \sum_\alpha  \mu_\alpha^2 }  \:  | \beta | ,  \;  \;
   \frac{ | \beta |  \sum_\alpha  \mu_\alpha^2 }{J}
   \ll  1 .
\end{align}
More-precise forms for the constraints depend on boundary conditions
and appear in App.~\ref{sec_Calc_Mus}.
The inverse temperature evaluates to
\begin{align}
   \label{eq_Beta_Predict}  
   \beta  
   =  \frac{ - E^\tot }{ 3 (2 \Sites \sites - 3) J^2 }
   + O_2   \, .
\end{align}
$O_2$ stands for 
``terms of second order in the small parameters in~\eqref{eq_High_T}.''
The effective chemical potentials evaluate to
\begin{align}
   \label{eq_Mu_Predict}  
   \mu_\alpha
   =  - \frac{ 3 (2 \Sites \sites - 3) }{ \Sites \sites }  \:
         \frac{ S_\alpha J^2 }{ E^\tot }
   +  O_2   \, .
\end{align}

In the thermodynamic limit, 
$H^\inter$ drops out of the prediction~\eqref{eq_O_Predict_Intro},
as discussed in the introduction.
If all $\Sys$ observables $O$ have NATS expectation values,
$\Sys$ thermalizes to the NATS state~\eqref{eq_NATS}.
Outside the thermodynamic limit,
noncommutation may prevent $\rho_\Sys$
from reaching $\rho_\NATS$ precisely~\cite{NYH_16_Microcanonical}.
The distance between the states
was quantified with the relative entropy,
\begin{align}
   \label{eq_Rel_Ent_Def}
   D ( \rho_\Sys || \rho_\NATS )
   =  \log \LParen \rho_\Sys 
   [ \log  \rho_\Sys  -  \log \rho_\NATS ]  \RParen .
\end{align}
Logarithms are base-$e$ throughout this paper.
The relative entropy quantifies the accuracy with which 
$\rho_\Sys$ can be distinguished from $\rho_\NATS$,
on average, in a binary hypothesis test~\cite{NielsenC10}.
The relative entropy~\eqref{eq_Rel_Ent_Def} was predicted to decline as 
the number $\Sites$ of systems grows~\cite{NYH_16_Microcanonical}:
\begin{align}
   \label{eq_Dist_Result}
   D \left( \rho_\Sys  ||  \rho_\NATS  \right)
   & \leq  \frac{ \const }{ \sqrt{\Sites} }  +  \const
\end{align}
This scaling can be checked with quantum state tomography~\cite{Banaszek_13_Focus}
in the finite-size experiments feasible today.
We detail the tomographic process 
in App.~\ref{sec_Tomography}.
Numerical simulations point to a scaling close to~\eqref{eq_Dist_Result}
(Sec.~\ref{sec_Numerics}).
The constant term in~\eqref{eq_Dist_Result} 
comes from the charges' noncommutation.
The constant depends on 
the parameters that quantify how much 
the definition of ``microcanonical subspace''
is relaxed to include $\amc$.
The larger the whole system,
the better the $(Q_\alpha^\tot / \Sites)$'s commute,
so the less the definition needs relaxing,
so the greater the probability that
some $\amc$ corresponds to a smaller constant.

\section{Numerical simulations}
\label{sec_Numerics}


We numerically simulated the experimental protocol
via direct calculation.
The spin chain's length varied from $\Sites \sites = 6$ to 14 qubits.
The first two qubits served as $\Sys$,
without loss of generality due to periodic boundary conditions.

We followed the first state preparation protocol in Sec.~\ref{sec_Proposal}:
The first six qubits were prepared in
$\ket{ x+ }  \ket{ z+ }  \ket{ x- }  \ket{ z- }  \ket{ x- }  \ket{ z+ }$;
and the rest of the qubits, in copies of $\ket{ z- }  \ket{ z+ }$.
Hence the total charges had the expectation values
$S_x = -1$, $S_y = 0$, and $S_z = 1$.

The state evolved under the Hamiltonian~\eqref{eq_HTot}
for a time $t = 2^{\Sites \sites}$,
wherein $J = 1$.
The exponential time sharpens the distinction between
the NATS and canonical predictions.
However, a time $t \sim \Sites \sites$ suffices.
Usually, when simulating charge-conserving evolution,
one represents the Hamiltonian as a matrix
relative to an eigenbasis shared with the charges.
The matrix is block-diagonal, simplifying calculations.
Here, the charges $\sigma_\alpha^\tot$ share no eigenbasis,
due to their noncommutation.
Hence $H^\tot$ does not block-diagonalize 
in terms of an eigenbasis shared by the charges,
and calculations do not simplify accordingly.
Relatedly, calculating the NATS's $\beta$ and $\mu_\alpha$'s from 
Eqs.~\eqref{eq_Beta_Predict_Intro} and~\eqref{eq_Mu_Predict_Intro}
numerically would cost considerable computation.
Four matrix-containing equations must be solved from four unknowns.
Hence the parameters were calculated analytically,
from analogs of Eqs.~\eqref{eq_Beta_Predict} and~\eqref{eq_Mu_Predict}
that follow from periodic boundary conditions 
(App.~\ref{sec_Calc_Mus}).
Using the analytics requires us to simulate
high, though finite temperatures 
and low, though nonzero, $\mu_\alpha$'s.
Extensions to $|T| \gtrsim 0$ and $|\mu_\alpha| \gg 0$
may be facilitated by, e.g., techniques in~\cite{Alhassid_78_Upper,Agmon_79_Algorithm}.
The calculations are correct to first order 
in the small parameters in Eq.~\eqref{eq_High_T}.

%
\begin{figure}[hbt]
\centering
\includegraphics[width=.5\textwidth, clip=true]{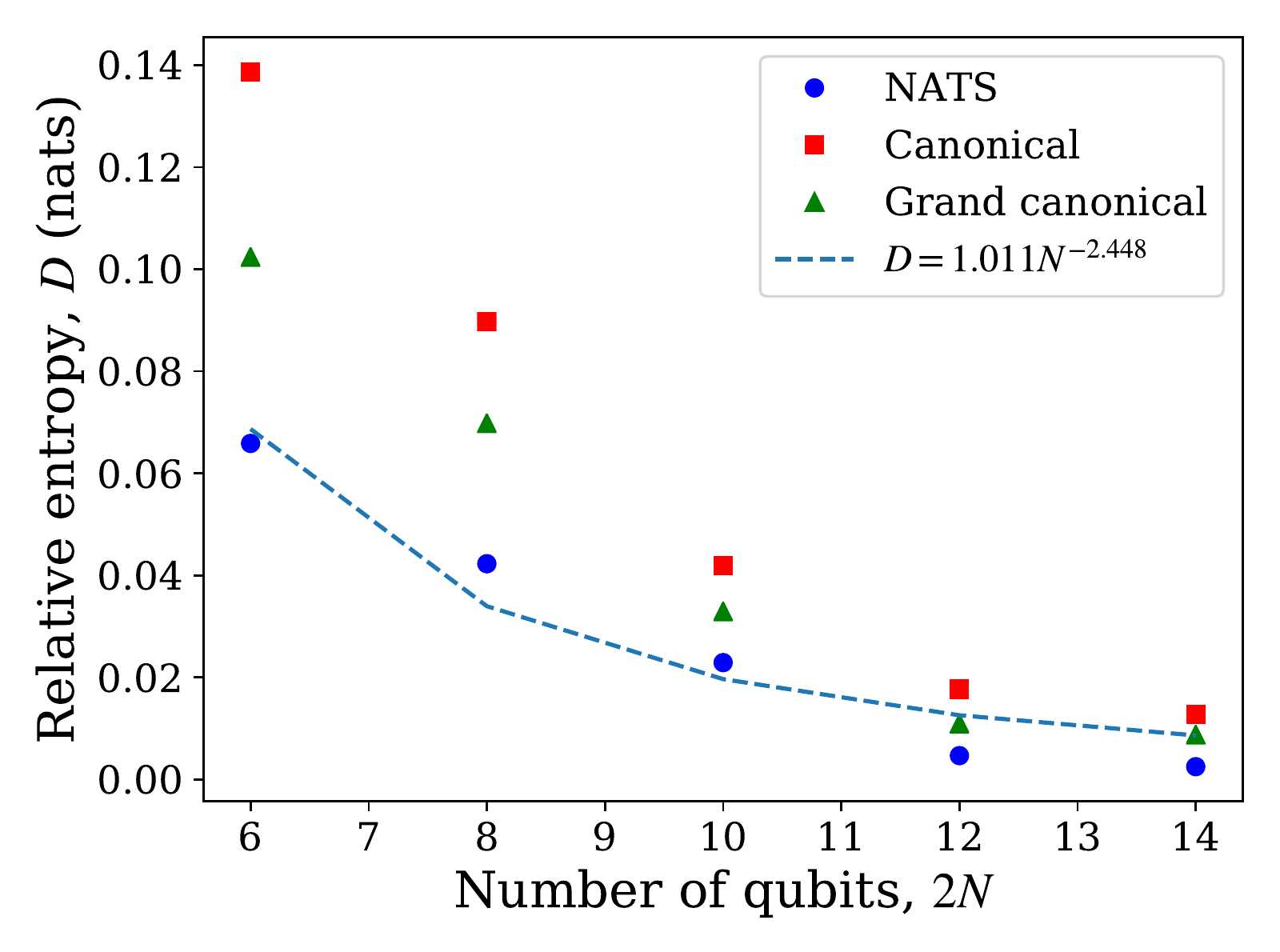}
\caption{\caphead{Distances from the long-time state 
to thermal states:} 
A chain of qubits subject to periodic boundary conditions
was simulated numerically.
The first preparation protocol and the evolution 
in Sec.~\ref{sec_Proposal} thermalized
a two-qubit system $\Sys$ of interest
to a long-time state $\rho_\Sys$.
Plotted is the relative-entropy distance [Eq.~\eqref{eq_Rel_Ent_Def}]
from $\rho_\Sys$ to each of three thermal states: 
the NATS [Eq.~\eqref{eq_NATS}] (blue dots),
the canonical state (red squares), 
and a grand canonical state 
$\propto  e^{ - \beta (H^\Sys - \mu_z \sigma_z^\Sys) }$
(green triangles).
The NATS theory predicts $\rho_\Sys$ with greater accuracy,
which grows with the spin-chain size, for finite systems.
The dashed line represents the best polynomial fit.
Entropies are expressed in units of nats
(not to be confused with the NATS:
logarithms are base-$e$).
}
\label{fig_Numerics}
\end{figure}

The final system-of-interest state $\rho_\Sys$ was compared to 
the NATS prediction~\eqref{eq_NATS},
to the canonical prediction~\eqref{eq_Rho_Can},
and to a grand canonical state
$\rho_\GC  \propto  e^{ - \beta (H^\Sys - \mu_z \sigma_z^\Sys) }$
that follows from ignoring the conservation of 
two noncommuting charges.
The canonical and grand canonical comparisons were modeled on 
the comparison to microcanonical and grand canonical predictions
in the original GGE studies~\cite{Rigol_07_Relaxation,Rigol_09_Breakdown}.
The canonical state's $\beta$ equals the NATS's to first order
in the dimensionless parameters [Eq.~\eqref{eq_High_T}]. 
The more-precise versions of inequalities~\eqref{eq_High_T} 
(App.~\ref{sec_Calc_Mus}) were satisfied:
Each small parameter was an order of magnitude less than 1.\footnote{
One exception arises at small system sizes:
When $\Sites \sites = 6, 8$, 
two small parameters equal $0.667, 0.500 < 1$.}

Figure~\ref{fig_Numerics} shows the relative entropies.
The blue dots show $D ( \rho_\Sys || \rho_\NATS )$;
the red squares, $D ( \rho_\Sys || \rho_\can )$;
and the green triangles, $D ( \rho_\Sys || \rho_\GC )$.
The NATS theory predicts $\rho_\Sys$ with greater accuracy,
which grows with the spin chain.

The dashed line shows the best polynomial fit,
which scales as approximately $\Sites^{-5/2}$.
This fit merits comparison with the right-hand side of 
Eq.~\eqref{eq_Dist_Result}.
As the system size $\Sites$ grows, the best fit shrinks more quickly
than the $\sim \Sites^{-1/2}$ prediction in~\cite{NYH_16_Microcanonical}.
This contrast suggests two possibilities:
(i) A bound tighter than that in~\cite{NYH_16_Microcanonical}
can be proved.
(ii) ``Transients,'' such as $\sim \Sites^{-5/2}$, 
dominate the scaling at small system sizes.
The transients vanish quickly as $\Sites$ grows, 
and $\sim \Sites^{-1/2}$ dominates the scaling at large system sizes.

As the whole system grows,
the NATS, grand canonical, and canonical predictions
appear to converge.
On the right-hand side of Fig.~\ref{fig_Numerics},
the blue circle, green triangle, and red square clump close together.
This converge is believed to result from 
the largeness of $T$ and the smallness of the $\mu_\alpha$'s.
When the temperature is high,
all thermal ensembles resemble the maximally mixed state,
$\id / 2^{\Sites \sites}$.
The predictions are expected to separate 
as $T$ falls and the $\mu_\alpha$'s grow.

In experiments, a Hamiltonian close to have the Heisenberg form
$\sum_\alpha \sigma_\alpha^\JParen  \sigma_\alpha^{(j+1)}$
can suffer from anisotropies~\cite{Fukuhara_13_Microscopic}.
Appendix~\ref{sec_Robust} demonstrates 
our protocol's robustness with respect to realistic anisotropies.

\section{Discussion}
\label{sec_Discussion}

We have formulated and simulated an experimental protocol 
for thermalizing a quantum many-body system to the NATS.
The protocol holds promise for ultracold atoms, trapped ions, quantum dots, 
nitrogen-vacancy centers, and NMR.
This work initiates a bridge from 
the abstract, idealized NATS theory of QI-theoretic thermodynamics
to many-body physics:
We introduce noncommutation---a key feature of nonclassicality---of charges
into condensed matter and AMO physics.
Extensions to high-energy physics beg to be realized.
Below, we contrast the NATS with the GGE.
Similarly, in App.~\ref{sec_U1}, we detail how noncommuting charges
invalidate predictions by the ETH.
However, Deutsch's original argument for studying the ETH
provides extra motivation for studying the NATS (App.~\ref{app_Deutsch}).
Then, we present opportunities for future research.

The GGE is an ensemble to which quantum systems equilibrate
if extensively many nontrivial charges are conserved~\cite{Rigol_09_Breakdown,Rigol_07_Relaxation,Vidmar_16_Generalized}.
Our prediction lies outside existing GGE studies for three reasons.
First, GGE studies have not emphasized noncommutation
(though noncommuting charges have now appeared in~\cite{Fagotti_14_On}).
Second, the GGE was designed for integrable Hamiltonians.
Our Hamiltonian is nonintegrable, 
because we study thermalization.
Third, the charges conserved in GGE problems 
tend not to equal sums of local charges.
Our globally conserved charges do, in the spirit of
the textbook problem reviewed in the introduction.
We maintain this spirit to emphasize that
beginning with a textbook problem
and introducing the minimal noncommutative tweak
unmoors conventional expectations,
as explained in the paragraph above Eq.~\eqref{eq_NATS}.
Our work moors this nonclassical thermalization 
to an experimental protocol and numerical simulations.

This paper opens up several opportunities for future research.
In condensed-matter, AMO, and high-energy physics
have recently emerged 
toolkits for studying many-body thermalization: 
quantum-simulator experiments~\cite{Neill_16_Ergodic,Smith_16_Many,Bernien_17_Probing,Wei_18_Emergent,Kucsko_18_Critical}, 
the ETH~\cite{Deutsch_91_Quantum,Srednicki_94_Chaos,Rigol_08_Thermalization},
random unitary circuits~\cite{Brown_12_Scrambling,Nahum_18_Operator,Khemani_18_Operator,HunterJones_18_Operator},
the GGE~\cite{Rigol_07_Relaxation,Rigol_09_Breakdown,Langen_15_Experimental},
and out-of-time-ordered correlators~\cite{Swingle_18_Quantitative}.
These toolkits should and can be generalized 
to accommodate noncommuting charges,
now that such charges have been imported 
from QI-theoretic thermodynamics into many-body physics.

Furthermore, these frameworks can be leveraged to explore 
noncommutation's effects on thermalization.
Constraining dynamics, noncommutation might slow 
the transport of energy, information, and/or charges.
Hence noncommutation might enhance storage and memory.
Additionally, noncommutation underlies 
quantum error correction, quantum cryptography, and other applications.
Noncommutation might advance information processing in materials.
Furthermore, group theory structures high-energy physics.
Non-Abelian groups therein might give rise to NATS physics.

The thermodynamic limit, too, merits study.
We focus on experimentally realizable systems, of finite size $\Sites$.
Figure~\ref{fig_Numerics} suggests that,
as the whole system grows, 
the canonical prediction's accuracy grows.
How much the NATS prediction outperforms the canonical
as $\Sites \to \infty$ 
remains an open question.

Degeneracies suggest more questions:
The ETH elucidates how quantum many-body systems thermalize
under nondegenerate Hamiltonians.
Conserved charges introduce degeneracies,
which can affect thermodynamic ensembles.
We address degeneracy through
the microcanonical lens of~\cite{NYH_16_Microcanonical}:
Noncommutation can prevent the charges from sharing an eigenspace.
No degenerate microcanonical subspace necessarily exists.
The microcanonical subspace was therefore generalized to
the a.m.c. subspace~\cite{NYH_16_Microcanonical}.
We have proposed protocols for preparing a global system
in an a.m.c. subspace.
This QI-thermodynamic approach to degeneracy
should be complemented with
a many-body-physics approach.

%
%
\begin{acknowledgments}
The authors are grateful to many people for illuminating discussions: \'Avaro Mart\'in Alhambra, Yoram Alhassid, Antoine Browaeys, Lincoln Carr, Vanja Dunjko, Manuel Endres, Philippe Faist, Markus Greiner, Vedika Khemani, Julian L\'{e}onard, Seth Lloyd, Mikhail Lukin, Noah Lupu-Gladstein, Maxim Olshanyi, Jonathan Oppenheim, Asier Pi\~{n}eiro Orioli, Ana Maria Rey, Marcos Rigol, Vladan Vuletic, Andreas Winter, and Mischa Woods.
NYH is grateful for funding from the Institute for Quantum Information and Matter, an NSF Physics Frontiers Center (NSF Grant PHY-1125565) 
with support from the Gordon and Betty Moore Foundation (GBMF-2644);
for an NSF grant for the Institute for Theoretical Atomic, Molecular, and Optical Physics at Harvard University and the Smithsonian Astrophysical Observatory;
and for hospitality at the KITP (supported by NSF Grant No. NSF PHY-1748958),
during its 2018 ``Quantum Thermodynamics'' conference.
AK acknowledges support from the US Department of Defense.
\end{acknowledgments}

\begin{appendices}

\onecolumngrid

\renewcommand{\thesection}{\Alph{section}}
\renewcommand{\thesubsection}{\Alph{section} \arabic{subsection}}
\renewcommand{\thesubsubsection}{\Alph{section} \arabic{subsection} \roman{subsubsection}}

\makeatletter\@addtoreset{equation}{section}
\def\theequation{\thesection\arabic{equation}}

\section{Protocol's inequivalence to
thermalization under a Hamiltonian that has only U(1) symmetry}
\label{sec_U1}

One might worry that our protocol's final state
could be predicted without the NATS theory,
that the NATS adds nothing to our knowledge of thermalization.
\emph{Prima facie}, the prediction seems to require 
knowledge of only the ETH 
and thermalization under U(1)-symmetric Hamiltonians.
The latter thermalization has been studied in,
e.g.,~\cite{Khemani_18_Operator,HunterJones_18_Operator}.
A U(1)-symmetric qubit Hamiltonian conserves $\sigma_z^\tot$.
The Hamiltonian is equivalent,
via a Jordan-Wigner transformation,
to a Hamiltonian that conserves particle number.
Hence systems thermalize to the grand canonical ensemble
under U(1)-symmetric evolution.
This thermalization, we show, is inequivalent to
our protocol's thermalization:
Justifiably predicting our protocol's final state
requires knowledge of the NATS.
Afterward, we present three more reasons for
the inequivalence of thermalization to the NATS and
thermalization under a U(1)-symmetric Hamiltonian:
First, microscopic dynamics distinguish the two thermalization processes.
Second, thermalization to the NATS is inequivalent to 
thermalization to the grand canonical ensemble
just as thermalization to the grand canonical state is inequivalent to 
thermalization to the canonical state.
Third, our thermalization protocol and 
thermalization to the grand canonical state
lead to thermal states
whose group-theoretic properties differ.

A brief review of the ETH is in order~\cite{Deutsch_91_Quantum,Srednicki_94_Chaos,Rigol_08_Thermalization,D'Alessio_16_From}.
The ETH governs a chaotic quantum many-body system
evolving under a nondegenerate Hamiltonian
$H^\tot = \sum_m  E_m  \ketbra{m}{m}$.
Suppose that $H^\tot$ conserves no nontrivial charges.
Let $O$ denote a local observable.
A matrix with elements $O_{m n}$ represents $O$
relative to the energy eigenbasis.
The diagonal elements $O_{mm}$ vary little with $m$, according to the ETH.
Furthermore, off-diagonal elements $O_{m (n \neq m)}$
are exponentially small in the system size.
The ETH implies ergodicity, thermalization to a microcanonical
(or canonical) expectation value~\cite{Rigol_12_Alternatives}.
In many studies, $H^\tot$ conserves a charge, such as $\sigma_z^\tot$.
The ETH is justified within a charge sector.

First, we elucidate why our protocol's final state
appears predictable with just
knowledge of the NATS and 
of thermalization under U(1)-symmetric Hamiltonians.
Imagine learning our protocol's initial state, $\rho$,
and Hamiltonian, $H^\tot$.
Imagine having to predict the final state's form
without knowing the NATS theory.
One might reason as follows:
$H^\tot$ has SU(2) symmetry.
$\sigma_x^\tot$, $\sigma_y^\tot$ and $\sigma_z^\tot$
generate SU(2).
Hence the evolution conserves
$\expval{  \sigma_\alpha^\tot  }
=  \Tr  \left( \rho  \sigma_\alpha^\tot  \right)$
for all $\alpha = x, y, z$.
The expectation values form a vector
$( \expval{ \sigma_x^\tot } ,  \expval{ \sigma_y^\tot },  \expval{ \sigma_z^\tot } )
\equiv r \hat{r}$.
The coordinate system can be transformed
such that $\hat{r}$ coincides with the new $z$-direction, $\hat{z'}$.
The transformation conserves $H^\tot$.
In this reference frame, only $\expval{ \sigma_{z'}^\tot }  \neq  0$.
Furthermore, $\expval{ \sigma_{z'}^\tot (t) }$ remains constant.
The thermalization therefore appears, \emph{prima facie}, identical to
thermalization under a U(1)-symmetric Hamiltonian.
One might therefore predict that
the system of interest thermalizes to
a grand canonical state in this reference frame.
Knowing the ETH, one might predict Eq.~\eqref{eq_O_Predict_Intro},
wherein $\sum_\alpha \mu_\alpha  Q_\alpha
=  \mu_{z'}  \sigma_{z'}^\tot$,
despite misrepresenting the microscopic dynamics (see below).
One could extrapolate the ETH 
to reconstruct the NATS prediction.
Without the NATS theory, however,
this prediction would have even less justification
than most ETH claims.
(The ETH remains a hypothesis.
Analytical support for the ETH remains under construction.)

The ETH implies thermalization when the initial state's support
lies on a small microcanonical window of energy levels.
Consider the extension of the ETH to the grand canonical ensemble.
The Hamiltonian shares an eigenbasis with the particle-number operator.
The extension is justified when
the initial state's weight lies on
a small microcanonical window of shared eigenstates.
Now, consider extending the ETH to thermalization under 
a Hamiltonian that conserves
noncommuting charges $Q_\alpha^\tot$.
One would na\"ively expect the extension to be justified
when the initial state's support lies on
a small microcanonical window
of eigenstates shared by $H^\tot$ and all the $Q_\alpha^\tot$'s.
Earlier studies of thermalization in the presence of U(1) symmetry
would support the extension.
But the $Q_\alpha^\tot$'s do not necessarily share eigenstates,
as they fail to commute.
Hence an extension of the ETH seems impossible to justify\ldots unless
the notion of a microcanonical subspace is generalized
to an approximate microcanonical subspace.
This generalization forms a cornerstone of 
the NATS theory~\cite{NYH_16_Microcanonical}.
Hence the NATS theory is necessary
for justifiably predicting the state to which our system thermalizes.
This paper shows that the prediction is accurate
for finite-size spin chains evolving under Eq.~\eqref{eq_HTot}.

NATS thermalization is inequivalent to 
thermalization under a U(1)-symmetric Hamiltonian 
for three more reasons.
First, under U(1) symmetry, just two quantities hop between subsystems: 
energy and 
quanta of one component of angular momentum.
Quanta of all three components of the angular momentum---charges
that fail to commute with each other---hop
during thermalization to the NATS.
One misrepresents the microscopic dynamics
when attempting to reduce NATS thermalization
to thermalization under a U(1)-symmetric Hamiltonian.

The attempt's failure parallels the failure 
to reduce grand canonical thermalization to canonical thermalization.
The grand canonical state is
$\propto  e^{ - \beta (H - \mu N) }$,
wherein $H$ denotes a Hamiltonian, 
$N$ denotes a particle-number operator,
and $\mu$ denotes a chemical potential.
One can define an effective Hamiltonian
$\tilde{H} := H - \mu N$.
The grand canonical state will look identical to a canonical state,
$\propto e^{ - \beta \tilde{H} }$.
But this definition cannot reduce grand canonical physics
to canonical physics.
During thermalization to the canonical state,
subsystems exchange only energy.
During thermalization to the grand canonical state,
subsystems exchange energy and particles.
The very existence of the name ``grand canonical''
implies that the energy-and-particle problem
differs significantly from the canonical problem
and deserves independent consideration.
Analogously, one can redefine the $z$-axis
such that the NATS state in~\eqref{eq_O_Predict_Intro}
looks identical to the grand canonical ensemble.
But this redefinition cannot reduce NATS thermalization
to grand canonical,
just as a definition cannot reduce grand canonical
to canonical.

Finally, if the Hamiltonian has only U(1) symmetry, 
the thermal state is proportional to 
an exponential that contains a Hamiltonian that has only U(1) symmetry. 
The NATS contains a Hamiltonian that has a non-Abelian symmetry. 
The two states have different group-theoretic properties.

\section{In every short-range-correlated state,
each total spin component has 
a subextensive standard deviation.}
\label{sec_Variance}

Consider an arbitrary short-range-correlated state
of correlation length $\xi$.
Let $\expval{O}$ denote the expectation value of
an observable $O$ in that state.
By assumption,
\begin{align}
   \label{eq_Short_Range}
   \expval{  \sigma_\alpha^\JParen  \sigma_\alpha^{(j')}  }
   - \expval{ \sigma_\alpha^\JParen }
      \expval{  \sigma_\alpha^{(j')} }
   \sim  e^{ - | j - j' | / \xi } \, .
\end{align}
We have set the lattice spacing to one.
Let us calculate each term in the standard deviation of 
$\sigma_\alpha^\tot$,
\begin{align}
   \label{eq_Std_devn_scaling_1}
   \sqrt{  \expval{ ( \sigma_\alpha^\tot )^2 }
             -  \expval{  \sigma_\alpha^\tot  }^2  }  \, .
\end{align}

The first term has the form
\begin{align}
   \expval{ ( \sigma_\alpha^\tot )^2 }
   & =  \expval{  \left(  
   \sum_{j = 1}^{\Sites \sites}  \sigma_\alpha^\JParen  \right)
   \left(  \sum_{j' = 1}^{\Sites \sites}  \sigma_\alpha^{(j')}  \right)  } 
   \label{eq_Std_Devn_Help1}
   =  \expval{  \sum_{j = 1}^{\Sites \sites}  ( \sigma_\alpha^\JParen )^2  }
   +  \sum_{j \neq j'}  \expval{  \sigma_\alpha^\JParen 
                                                \sigma_\alpha^{(j')}  }  .
\end{align}
The first term on the right-hand side of Eq.~\eqref{eq_Std_Devn_Help1} simplifies as
\begin{align}
   \label{eq_Std_Devn_Help1b}
   \expval{  \sum_{j = 1}^{\Sites \sites}  ( \sigma_\alpha^\JParen )^2  }
   & =  \sum_{j = 1}^{\Sites \sites}  
   \expval{  \id_2^{\otimes \Sites \sites}  } 
   =  \Sites \sites  .
\end{align}
The second term on the right-hand side of Eq.~\eqref{eq_Std_Devn_Help1} 
simplifies under assumption~\eqref{eq_Short_Range}:
\begin{align}
   \label{eq_Std_Devn_Help2}
   \sum_{j \neq j'}  \expval{  \sigma_\alpha^\JParen 
                                                \sigma_\alpha^{(j')}  }
   \sim  \sum_{j \neq j'}     \expval{  \sigma_\alpha^\JParen  }
   \expval{  \sigma_\alpha^{(j')}  }
   +  \sum_{j \neq j'}   e^{ - | j - j' | / \xi }  .
\end{align}
The second term has significant contributions
only from subterms in which $j'$ lies within $\xi$ of $j$.
Hence the $e^{ - | j - j' | / \xi }  \sim  e^{ - \xi / \xi }  =  \const$
A constant number of such subterms exist.
Hence the right-hand side of Eq.~\eqref{eq_Std_Devn_Help2}
can be approximated with
\begin{align}
   \label{eq_Std_Devn_Help3}
   \sum_{j \neq j'}  \expval{  \sigma_\alpha^\JParen 
                                                \sigma_\alpha^{(j')}  }
   \sim  \sum_{j \neq k}     \expval{  \sigma_\alpha^\JParen  }
   \expval{  \sigma_\alpha^{(j')}  }
   +  \Sites  \sites  .
\end{align}
Substituting from Eqs.~\eqref{eq_Std_Devn_Help1b}
and~\eqref{eq_Std_Devn_Help3} into
the right-hand side of Eq.~\eqref{eq_Std_Devn_Help1} yields
\begin{align}  
   \label{eq_Std_Devn_Help4}
   \expval{ ( \sigma_\alpha^\tot )^2 }
   & \sim  \Sites \sites
   +  \sum_{j \neq j'}     \expval{  \sigma_\alpha^\JParen  }
   \expval{  \sigma_\alpha^{(j')}  }  .
\end{align}

Let us estimate the second term in~\eqref{eq_Std_devn_scaling_1}:
\begin{align}
   \expval{  \sigma_\alpha^\tot  }^2
   & =  \expval{  \sum_{j = 1}^{\Sites \sites}  
                         \sigma_\alpha^\JParen  }^2
   =  \left(  \sum_{j = 1}^{\Sites \sites}
               \expval{  \sigma_\alpha^\JParen  }
      \right)^2
   =  \left(  \sum_{j = 1}^{\Sites \sites}
                \expval{  \sigma_\alpha^\JParen  }  \right)
   \left(  \sum_{j' = 1}^{\Sites \sites}
                \expval{  \sigma_\alpha^{(j')}  }  \right) \\
   & =  \sum_{j = 1}^{\Sites \sites}
   \expval{  \sigma_\alpha^\JParen  }^2
   +  \sum_{j \neq j'}  
   \expval{  \sigma_\alpha^\JParen  }
   \expval{  \sigma_\alpha^{(j')}  }  \\
   \label{eq_Std_Devn_Help5}
   & \sim  \Sites \sites  
    +  \sum_{j \neq j'}  
   \expval{  \sigma_\alpha^\JParen  }
   \expval{  \sigma_\alpha^{(j')}  } .
\end{align}
We have approximated the first term with
$(\text{number of terms})(\text{operator norm of } \sigma_\alpha^\JParen)$.
Let us substitute from Eqs.~\eqref{eq_Std_Devn_Help4}
and~\eqref{eq_Std_Devn_Help5}
into Eq.~\eqref{eq_Std_devn_scaling_1}.
The $\sum_{j \neq k}$ terms cancel exactly, leaving
\begin{align}
   \label{eq_Std_devn_scaling_2}
   \sqrt{  \expval{ ( \sigma_\alpha^\tot )^2 }
             -  \expval{  \sigma_\alpha^\tot  }^2  }
   \sim  \sqrt{\Sites \sites} .
\end{align}

\section{Soft measurement}
\label{sec_Soft_Meas}

This appendix details the soft measurements introduced in Sec.~\ref{sec_Proposal}.
We formalize soft measurements in App.~\ref{sec_Formalize_Soft}.
Appendix~\ref{sec_Meas_Intuition} provides physical intuition
about the preparation procedure that relies on soft measurements.

\subsection{Formalization of soft measurements}
\label{sec_Formalize_Soft}

We formalize soft measurements with 
a positive operator-valued measure (POVM).
POVMs model generalized measurements in QI theory~\cite{NielsenC10}.
A POVM consists of positive operators $M_\ell > 0$,
called \emph{Kraus operators.}
They satisfy the completeness relation
$\sum_\ell M_\ell^\dag  M_\ell  =  \id$.
Measuring the $\Set{ M_\ell }$ of a state $\rho$
has a probability $\Tr ( M_\ell^\dag M_\ell \rho )$ 
of yielding outcome $\ell$.
The measurement updates $\rho$ to
$M_\ell \rho M_\ell^\dag / 
\Tr ( M_\ell^\dag M_\ell \rho )$.
Let $P_\alpha^{S_\alpha}$ denote the projector onto
the eigenvalue-$S_\alpha$ eigenspace of
$\sigma_\alpha^\tot$.
A soft $\sigma_\alpha^\tot$ measurement has the form
$\{ M_\alpha^{S_\alpha} \}$.
The outcome $S_\alpha$ labels the Kraus operators,
\begin{align}
   \label{eq_Soft_POVM}
   M_\alpha^{S_\alpha} 
   =  \sum_{ \tilde{S}_\alpha = - \Sites \sites,  - \Sites \sites + 2, 
                    \ldots, \Sites \sites - 2,  \Sites \sites } 
   \sqrt{ f_{\Sites \sites} ( S_\alpha, \tilde{S}_\alpha ) }  \;
   P_\alpha^{  \tilde{S}_\alpha  } .
\end{align}
Outputting $S_\alpha$, the measurement projects 
the state a little onto each of the eigenspaces in superposition.
How much does the measurement project onto 
the eigenspace associated with some eigenvalue $\tilde{S}_\alpha$?
The amount depends on the amplitude
$f_{\Sites \sites} ( S_\alpha, \tilde{S}_\alpha )$.
The amplitude must maximize where $S_\alpha = \tilde{S}_\alpha$,
to satisfy the peaking requirement (Sec.~\ref{sec_Proposal}).
The binomial distribution suggests itself.
We present the distribution, then derive and analyze it:
\begin{align}
   \label{eq_POVM_Envelope}  
   f_{\Sites \sites} (  S_\alpha ,  \tilde{S}_\alpha  ) 
  & =  {\Sites \sites  \choose  
         \frac{1}{2} (\Sites \sites + S_\alpha) }
   \left[ \frac{1}{2}  
           \left( 1 +  \frac{ \tilde{S}_\alpha }{ \Sites \sites }  \right)
   \right]^{ \frac{1}{2} (\Sites \sites + S_\alpha) }
   \left[ \frac{1}{2}  
           \left( 1 -  \frac{ \tilde{S}_\alpha }{ \Sites \sites }  \right)
   \right]^{ \frac{1}{2} (\Sites \sites - S_\alpha) } \, .
\end{align}
We define $0^0 \equiv 1$.
Numerics confirm that 
the POVM~\eqref{eq_Soft_POVM} satisfies 
the mild-disturbance condition (ii) 
in Sec.~\ref{sec_Proposal}.

The envelope~\eqref{eq_POVM_Envelope}
is constructed as follows.
We semiclassically model each qubit 
as pointing upward or downward along the $\alpha$-axis.
We formulate the binomial probability that
an $(\Sites \sites)$-qubit chain 
has a magnetization $S_\alpha$,
if the average-over-trials magnetization equals $\tilde{S}_\alpha$.
Let $n_\uparrow$ and $n_\downarrow$ denote
the numbers of upward- and downward-pointing qubits
in some configuration.
Let $p_\uparrow$ denote the probability that a given qubit points upward
and $p_\downarrow$, the probability that the qubit points downward.
We must solve for each of these quantities in terms of
$S_\alpha$, $\tilde{S}_\alpha$, and $\Sites \sites$.
As $\Sites \sites  =  n_\uparrow  +  n_\downarrow$
and $S_\alpha  =  n_\uparrow  -  n_\downarrow$,
$n_\uparrow  =  \frac{1}{2} (\Sites \sites  +  S_\alpha)$, and
$n_\downarrow  =  \frac{1}{2} (\Sites \sites  -  S_\alpha)$.
On average, $\tilde{S}_\alpha
=  (p_\uparrow  -  p_\downarrow)  \Sites \sites$
qubits point upward.
By normalization, $p_\downarrow  =  1  -  p_\uparrow$.
Hence $p_\uparrow  =  \frac{1}{2} \left(   
1  +  \frac{ \tilde{S}_\alpha }{ \Sites \sites }  \right)$, and
$p_\downarrow  =  \frac{1}{2} \left(  
1  -  \frac{ \tilde{S}_\alpha }{ \Sites \sites }  \right)$.
The binomial function has the form
$f_{\Sites \sites} ( S_\alpha,  \tilde{S}_\alpha )
   = {\Sites \sites  \choose  n_\uparrow}
   ( p_\uparrow )^{n_\uparrow}
   ( p_\downarrow )^{ n_\downarrow } .$
Substituting in yields Eq.~\eqref{eq_POVM_Envelope}.

As $\Sites \sites \to \infty$, the binomial approaches a Gaussian.
The Gaussian has a mean of $\expval{ S_\alpha }  =  \tilde{S}_\alpha$
and a standard deviation of
\begin{align}
   \label{eq_Std_Devn_App}
   \Delta  =  \frac{1}{2} \sqrt{ \Sites \sites
   \left( 1  +  \frac{ \tilde{S}_\alpha }{ \Sites \sites } \right)
   \left( 1  -  \frac{ \tilde{S}_\alpha }{ \Sites \sites } \right)  }
   \sim  \sqrt{ \Sites \sites } .
\end{align}
Hence
\begin{align}
   \label{eq_Gauss_Limit}
   \lim_{\Sites \sites  \to  \infty}
   f_{\Sites \sites} (S_\alpha,  \tilde{S}_\alpha)
   =  \exp  \left(
   - \frac{ (S_\alpha  -  \tilde{S}_\alpha)^2 }{ 2 \Delta^2 }
   \right)  \Big/ \sqrt{ 2 \pi \Delta^2 }  \, .
\end{align}

\emph{Prima facie}, $S_\alpha$ and $\tilde{S}_\alpha$
appear to have been swapped relative to their natural roles:
$S_\alpha$ was defined as 
the ``expected'' $\sigma_\alpha^\tot$ value in Sec.~\ref{sec_Proposal}.
But $\tilde{S}_\alpha$ determines the mean spin
in Eq.~\eqref{eq_POVM_Envelope}.
This swap impacts the function's behavior little:
$f_{\Sites \sites} (S_\alpha,  \tilde{S}_\alpha)$ 
peaks at $S_\alpha = \tilde{S}_\alpha$.
The peak grows higher and narrower as $\Sites \sites$ grows.
As $\Sites \sites \to \infty$, 
the envelope approaches a Gaussian symmetric under 
$S_\alpha  \leftrightarrow  \tilde{S}_\alpha$
[Eq.~\eqref{eq_Gauss_Limit}].
Normalization motivates the swap:
The POVM~\eqref{eq_Soft_POVM} must satisfy 
the completeness condition
$\sum_{S_\alpha} ( M_\alpha^{S_\alpha} )^\dag 
M_\alpha^{S_\alpha}
=  \id$.
The POVM does because the envelope 
is normalized as
$\sum_{S_\alpha}  f_{\Sites \sites}  (S_\alpha,  \tilde{S}_\alpha)
= 1$.

\subsection{Physical intuition about the soft-measurement preparation procedure}
\label{sec_Meas_Intuition}

\begin{figure}[hbt]
\centering
\includegraphics[width=.35\textwidth, clip=true]{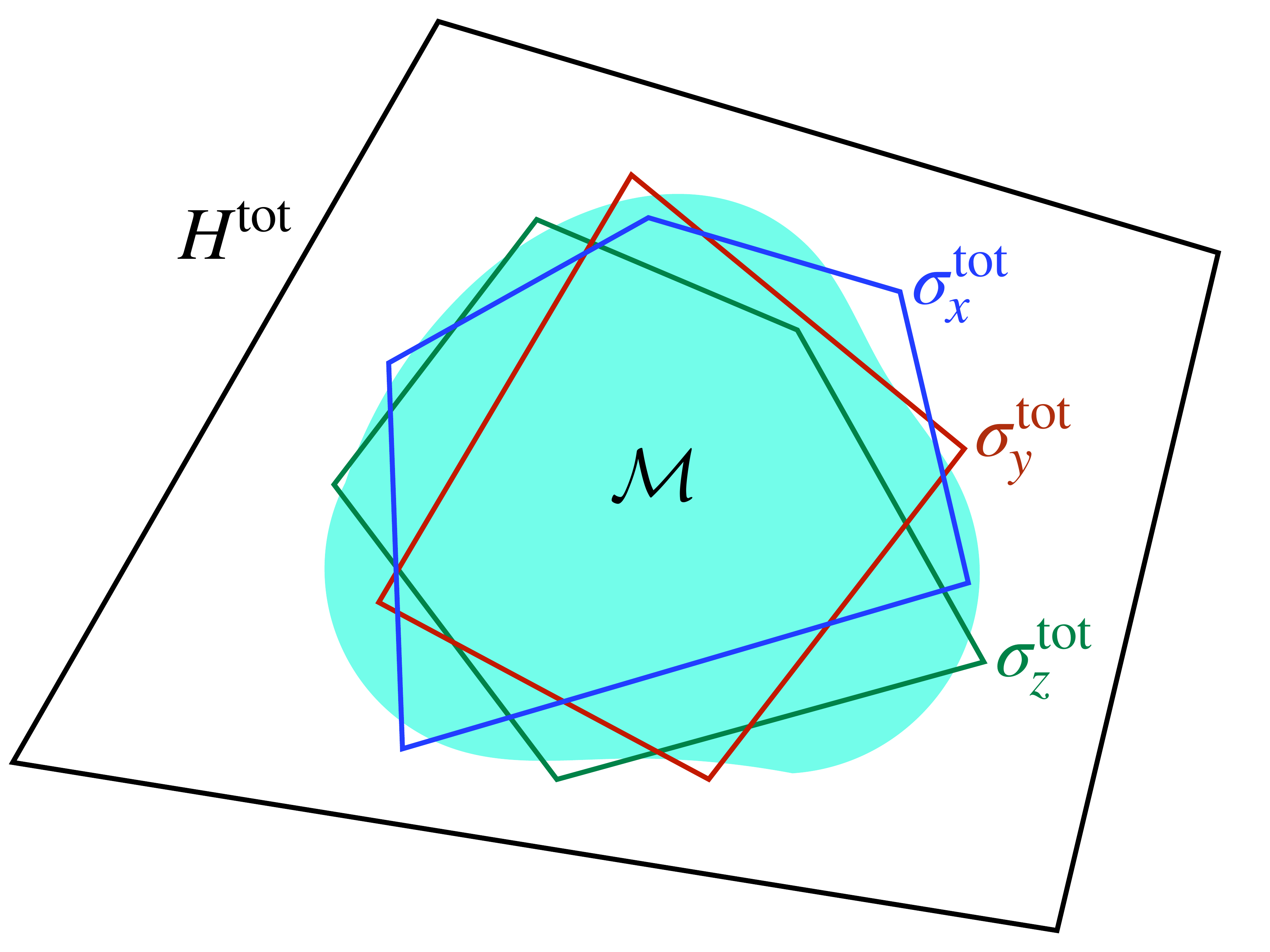}
\caption{\caphead{Sketch of subspaces:} 
The black, outermost line represents
an eigenspace of the total Hamiltonian, $H^\tot$.
Inside lies the approximate microcanonical subspace, $\amc$,
represented by the shaded shape.
$\amc$ generalizes the microcanonical subspace
to noncommuting exchanged charges.
The total-spin components $\sigma_{\alpha = x, y, z}^\tot$
have eigenspaces that largely coincide with $\amc$.}
\label{fig_amc_subspace}
\end{figure}

Suppose that the spin chain begins in a random state.
Measuring $H^\tot$ with decent precision 
projects the chain's state approximately onto
an energy eigenspace.
This eigenspace is larger than the a.m.c. subspace, $\amc$.
The soft $x$ measurement collapses the state a little,
shrinking the state's support.
The soft $y$ and $z$ measurements shrink the support further.
After the final measurement, at least most of the state's support
lies in $\amc$, as quantified in App.~\ref{sec_Params}.
Figure~\ref{fig_amc_subspace} sketches
the relationships amongst the subspaces.

Let us illustrate how each soft measurement 
partially collapses the spin chain's state.
Consider a toy system of $\Sites \sites = 2$ qubits
whose $\sigma_z^\tot$ and $\sigma_x^\tot$ are measured softly.
Suppose that the measurements yield $S_z, S_x = 0$.
The conditioned soft $z$ measurement projects the state with
$P_z^0  \propto  M_z^0$,
by Eqs.~\eqref{eq_Soft_POVM} and~\eqref{eq_POVM_Envelope}.
$P_z^0$ projects onto the eigenvalue-0 eigenspace of
$\sigma_z^\tot$.
This eigenspace is spanned by the singlet
$\ket{ s_z }  :=  \frac{1}{ \sqrt{2} } \:
( \ket{ z+, z-}  -  \ket{ z-, z+ } )$
and the entangled triplet
$\ket{ t_z }  :=  \frac{1}{ \sqrt{2} } \:
( \ket{ z+, z-}  +  \ket{ z-, z+ } )$.
That is,
$P_z^0  =  \ketbra{ s_z }{ s_z }  +  \ketbra{ t_z }{ t_z }$.
Similarly, the conditioned $x$ measurement projects the state with
$P_x^0  =  \ketbra{ s_x }{ s_x }  +  \ketbra{ t_x }{ t_x }$.

Onto what subspace does the sequence 
of approximate measurements project?
Let us express $P_x^0$ in terms of
the $z$-type singlet and triplets.
The singlet relative to any axis equals the singlet relative to every other,
to within a global phase:
$\ket{ s_x }  =  ( \text{phase} ) \ket{ s_z }$.
The $x$-type entangled triplet decomposes as
$\ket{ t_x }  \propto  \frac{ 1 }{ \sqrt{2} }  \:
(  \ket{ z+, z+ }  -  \ket{ z-, z- } )$.
Hence $P_x^0  P_z^0 =  \ketbra{ s_z }{ s_z }$.
The approximate $\sigma_z^\tot$ measurement
collapses the state onto a two-dimensional subspace;
and the approximate $\sigma_x^\tot$ measurement,
onto a one-dimensional subspace.

\section{Standard deviations in global charges $\sigma_\alpha^\tot$
after a sequence of soft measurements}
\label{app_Soft_Meas_Std_Devn}


Section~\ref{sec_Prep} and App.~\ref{sec_Soft_Meas}
introduce the soft-measurement protocol
for preparing a global state $\rho^\tot$ in an a.m.c. subspace.
$\rho^\tot$ should exhibit property~\ref{item_Std_Dev}:
In $\rho^\tot$, every global charge $\sigma_\alpha^\tot$
should have a standard deviation
that grows with the system size, $\Sites \sites$,
no more than linearly.
The need for slow scaling motivated 
soft measurements' ``mild-disturbance'' property.
Here, we numerically check the standard-deviation scaling 
at infinite temperature.

Setting $T = \infty$ obscures the distinction between 
the NATS and other thermodynamic ensembles
as measured in Sec.~\ref{sec_Numerics}.
However, this study offers two benefits:
First, these initial results motivate detailed numerics,
at large system sizes, outside the scope of this paper.
Second, $T = \infty$ will not necessarily hinder 
future thermodynamic studies of noncommuting charges.

\begin{figure}[hbt]
\centering
\includegraphics[width=.7\textwidth, clip=true]{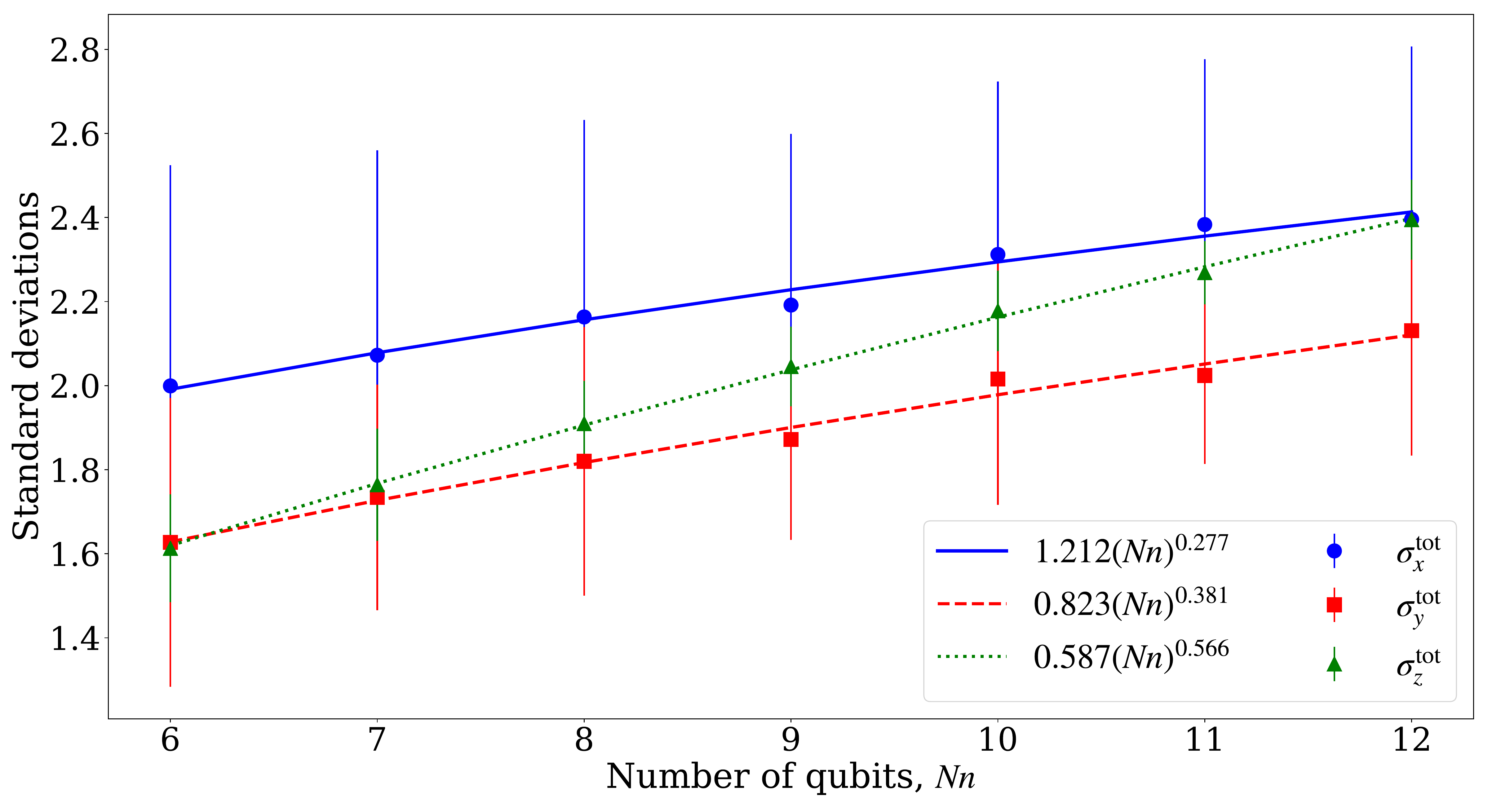}
\caption{\caphead{Standard deviations in global charges $\sigma_\alpha^\tot$
after a sequence of soft measurements:}
Each standard deviation has been averaged over 100 random initial global states,
which mimic infinite-temperature states.
The curves depict the best fits of the form $(\const) (\Sites \sites)^{\const}$.}
\label{fig_Soft_Meas_Std_Devn}
\end{figure}

%
%
%
We checked the standard deviations with the following protocol:
\begin{enumerate}

   \item \label{item_Std_Dev_Prep_State}
   Prepare the global system in a random pure initial state~\cite{Zyczkowski_11_Generating}: 
   Form a superposition of all the $\sigma_z$ product states.
   Choose each amplitude's real part according to 
   the standard normal distribution.
   Choose the amplitude's imaginary part, independently, 
   according to the same distribution.
   Normalize the state.
   
   \item
   Measure $\sigma_x^\tot$ softly.
   
   \item
   Measure $\sigma_y^\tot$ softly.
   
   \item 
   Measure $\sigma_z^\tot$ softly.
   
   \item \label{item_Std_Dev_Meas}
   Compute the standard deviation
   $\sqrt{ \expval{ ( \sigma_\alpha^\tot )^2 }  -  \expval{ \sigma_\alpha^\tot }^2  }$
   for every $\alpha$.
   
   \item 
   Perform steps~\ref{item_Std_Dev_Prep_State}--\ref{item_Std_Dev_Meas}
   for each of 100 random initial states.
   
   \item
   For each $\alpha$, average the standard deviation
   $\sqrt{ \expval{ ( \sigma_\alpha^\tot )^2 }  -  \expval{ \sigma_\alpha^\tot }^2  }$
   over the states.

\end{enumerate}
Figure~\ref{fig_Soft_Meas_Std_Devn} shows 
the state-averaged standard deviations
plotted against the global system size, $\Sites \sites$.
The curves show the best fits of the form $(\const) (\Sites \sites)^{\const}$.
If $\alpha = x, y$, the exponents are $0.277, 0.381 < 0.5$,
as required in property~\ref{item_Std_Dev} of Sec.~\ref{sec_Prep}.

If $\alpha = z$, the exponent lies slightly above $0.5$, at $0.566$.
The reason is, $\sigma_z^\tot$ was softly measured last:
In another study we softly measured $\sigma_z^\tot$,
then $\sigma_x^\tot$, then $\sigma_y^\tot$.
The last-measured charge scales with the greatest exponent, $0.612$.
However, the second-measured charge, $\sigma_x^\tot$,
scales with the least exponent.
In contrast, in Fig.~\ref{fig_Soft_Meas_Std_Devn}, the first-measured charge,
$\sigma_x^\tot$ scales with the least exponent.
We expect this discrepancy, as well as the slightly-above-$0.5$ exponent,
to disappear as the global system grows:
Large numbers promote internal averaging.

%
%
\section{Parameterization of the approximate microcanonical subspace}
\label{sec_Params}

An a.m.c. subspace $\amc$ is defined in terms
of five small parameters~\cite{NYH_16_Microcanonical}.
They govern the constants in Ineq.~\eqref{eq_Dist_Result},
the bound on the distance between $\rho_\Sys$ and the NATS.
The constants' forms are calculated partially in~\cite{NYH_16_Microcanonical}.
Calculating them completely would require experiments or extensive analytics.
We review the a.m.c. subspace's definition in Sec.~\ref{sec_Def_AMC}.
In Sec.~\ref{sec_Choose_Params},
we identify parameter values suited to our protocol.
We focus on the soft-measurement state preparation for concreteness.

\subsection{Definition of the a.m.c. subspace}
\label{sec_Def_AMC}

$\amc$ is defined in terms of two conditions~\cite{NYH_16_Microcanonical}:
(i) Every state in $\amc$ has
a fairly well-defined value of each $\sigma_\alpha^\tot$.
(ii) Consider state whose $\sigma_\alpha^\tot$
has a fairly well-defined value for every $\alpha$.
Most of the state's support lies in $\amc$.
These conditions are quantified in terms of small parameters
$\delta, \eta, \delta', \eta', \epsilon  \gtrsim  0$.

(i) Let $\omega$ denote any whole-system state
supported in just $\amc$.
In $\omega$, every total charge has a fairly well-defined value:
Consider measuring any $\sigma_\alpha^\tot$.
The measurement has a high probability
of yielding an outcome close to the ``expected value'' $S_\alpha$.
(The notation $v_\alpha = \frac{ S_\alpha }{ \Sites }$
is used in~\cite{NYH_16_Microcanonical}.)
Consider a narrow strip of $\sigma_\alpha^\tot$ eigenvalues 
centered on $S_\alpha$.
Recall that the system-of-interest charge 
$\sigma_\alpha^\JParen  +  \sigma_\alpha^{(j+1)}$ 
has a spectral diameter of two.
The strip is therefore chosen to extend a distance 
$2 \eta \Sites$ on either side of $S_\alpha$.
Consider the $\sigma_\alpha^\tot$ eigenvalues in
$[S_\alpha  -  2 \eta \Sites ,  \:
S_\alpha  +  2 \eta \Sites ]$.
They correspond to eigenspaces whose direct sum
is projected onto by $\Pi^\eta_\alpha$.
A $\sigma_\alpha^\tot$ measurement has a probability
$\Tr ( \omega  \Pi^\eta_\alpha )$
of yielding a value in this interval.
The probability must be at least $1 - \delta$:
\begin{align}
   \label{eq_AMC_Def1}
   \supp (\omega)  \subset  \amc
   \qquad \Rightarrow \qquad
   \Tr ( \omega  \Pi^\eta_\alpha )  \geq  1  -  \delta
   \quad  \forall \alpha .
\end{align}

(ii) Let $\omega'$ denote any state for which
measuring any $\sigma_\alpha^\tot$ has 
a high probability of yielding an outcome
close to the expected value,
within $2 \eta' \Sites \sites$ of $S_\alpha$.
Most of the support of $\omega'$ lies in $\amc$---at 
least a fraction $1 - \epsilon$.
As $P_\amc$ denotes the projector onto the a.m.c. subspace,
\begin{align}
   \label{eq_AMC_Def2}
   \Tr \left(  \omega'  \Pi^{\eta'}_\alpha  \right)
   \geq 1 - \delta'
   \quad  \forall \alpha
   \qquad  \Rightarrow  \qquad
   \Tr ( \omega'  P_\amc )
   \geq  1 - \epsilon .
\end{align}

\subsection{Parameter values suited to the soft-measurement preparation procedure}
\label{sec_Choose_Params}

We have freedom in choosing the parameters' values:
We have specified a procedure for preparing
a global state $\rho_\tot$ that has substantial support on
an a.m.c. subspace $\amc$.
In fact, $\rho_\tot$ might have much support on
each of multiple a.m.c. subspaces.
Hence we may be able to specify
one of multiple possible sets of parameter values.

Three principles guide our choice:
informativeness, tradeoffs, and the soft measurement's form.
First, some choices of parameters are less informative than others.
The greater the parameters, the less an a.m.c. subspace $\amc$
resembles a microcanonical subspace,
the less the global system is expected to thermalize internally,
and the looser the bound on $D( \rho_\Sys || \rho_\NATS )$
is expected to be.
Yet the parameters cannot be arbitrarily small,
because they trade off, second.
For example, the lesser the $\eta$,
the greater $\delta$ will tend to be,
by the inequality in~\eqref{eq_AMC_Def1}.
Third, the soft measurement's form points to
a natural choice of parameter values.
For example, $\approx 68\%$ of a Gaussian's support lies within
a standard deviation of its mean.
Hence we will choose $\delta' = 1 - 0. 68$.
This choice is not unique but is suggested by the soft measurement's form.

After the procedure, measuring any $\sigma_\alpha^\tot$
likely yields a value within a standard deviation $\Delta$
of $S_\alpha$.
The standard deviation scales as 
$\Delta  \sim  \sqrt{\Sites \sites}$.
Hence the procedure prepares an instance of
the $\omega'$ in~\eqref{eq_AMC_Def2},
for $2  \eta'  \Sites   =  (\const) \sqrt{\Sites \sites}$.
Rearranging yields the first small parameter,
\begin{align}
   \label{eq_Eta_Prime}
   \boxed{  \eta'
   =  (\const) / \sqrt{\Sites}  } \, .
\end{align}
We have incorporated $\sqrt{\sites}  =  \sqrt{2}$ into the constant.
The spin chain is large, so $\eta'$ is small, as desired.

We choose $\delta'$ by calculating
the left-hand side of the leftmost inequality in~\eqref{eq_AMC_Def2},
the probability that measuring 
a $\sigma_\alpha^\tot$ of $\omega'$
yields a value within $\Delta$ of $S_\alpha$.
We integrate $f_{\Sites \sites} (S_\alpha,  \tilde{S}_\alpha)$
across a region, centered on $\tilde{S}_\alpha = S_\alpha$,
of half-width $\Delta$:
\begin{align}
   \label{eq_Delta_Prime}
   1 - \delta'
   \leq  \int_{S_\alpha  -  \Delta }^{ S_\alpha  +  \Delta }
   d \tilde{S}_\alpha  \;
   f_{\Sites \sites}  (S_\alpha,  \tilde{S}_\alpha) .
\end{align}
We approximate $f_{\Sites \sites}$ with
the Gaussian~\eqref{eq_Gauss_Limit}.
A Gaussian is well-known to have $68 \%$ of its weight
within a standard deviation of its mean.
Hence we choose $1 - \delta' = 0.68$, or
\begin{align}
   \boxed{  \delta'  =  0.32 }  \, .
\end{align}
$\delta' \ll 1$, 
as desired.

We have chosen values for two of the five parameters 
that define an a.m.c. subspace $\amc$, $\eta'$ and $\delta'$.
Let us turn to $\eta$, $\delta$, and $\epsilon$.
In~\cite{NYH_16_Microcanonical},
$c$ denotes the number of non-Hamiltonian charges.
Theorem~4 in~\cite[Suppl. Inf.]{NYH_16_Microcanonical}
presents a condition under which $\amc$ is known to exist.
The condition governs the small parameters and 
the number $\Sites$ of subsystems:
For every $\epsilon > (c + 1) \delta' > 0$,
$\eta > \eta' > 0$, 
$\delta > 0$,
and all great-enough $\Sites$,
``there exists an $\left( \epsilon, \eta, \eta', \delta, \delta' \right)$-approximate
microcanonical subspace $\amc$ 
[\ldots] associated with [\ldots]
the approximate expectation values'' $S_\alpha$.
An $\amc$ might exist under other conditions.
But these known conditions motivate choices of
$\epsilon$ and $\eta$.
The theorem suggests choosing
\begin{align}
   \label{eq_Eta_Choice}
   \boxed{  \eta > \eta'  =  (\const) / \sqrt{\Sites} .}  \, .
\end{align}
By Eq.~\eqref{eq_Delta_Prime} and $c = 3$,
\begin{align}
   \boxed{
   \epsilon > (c + 1) \delta' = 1.28  }  \, ,
\end{align}
Though $\epsilon > 1$ contradicts the spirit
of the a.m.c. subspace's definition,
the inequality does not contradict the letter.

We have chosen values for all the parameters except $\delta$.
The soft-measurement procedure underdetermines $\delta$.
For inspiration for our choice of $\delta$,
we turn to the a.m.c. subspace's definition,~\eqref{eq_AMC_Def1} 
and~\eqref{eq_AMC_Def2}.
$\delta$ and $\delta'$ play analogous roles. 
Hence we choose 
\begin{align}
   \boxed{ \delta = \delta' = 0.32 }  \, .
\end{align}

%
%
%

%
%
%
\section{Calculation of the inverse temperature $\beta$
and the effective chemical potentials $\mu_\alpha$}
\label{sec_Calc_Mus}


Let us derive Eqs.~\eqref{eq_Beta_Predict} and~\eqref{eq_Mu_Predict}.
We index such that $\Sys$ consists of qubits $j = 1, 2, \ldots, \sites$.
This choice is for convenience,
and $\Sys$ lies far from the boundaries.
We assume that the temperature is high 
and the chemical potentials are low:
Calculations are to first order in
the small parameters approximated in the left-hand side of Ineq.~\eqref{eq_High_T}
and presented precisely below
[in Ineq.~\eqref{eq_High_T_App} for closed boundary conditions 
and in Ineq.~\eqref{eq_Small_Params_PBCs} for periodic].
We calculate the partition function, then $\beta$, and then the $\mu_\alpha$'s.
Rewriting Eq.~\eqref{eq_HTot} will prove convenient:
\begin{align}
   \label{eq_HTot_App}
   H^\tot
   =  J  \sum_{ \alpha = x, y, z }  \left(  
   \sum_{j = 1}^{\Sites \sites - 1}
   \sigma_{\alpha}^\JParen  \sigma_{ \alpha }^{(j + 1)}
   +  \sum_{j = 1}^{\Sites \sites - 2}
   \sigma_{\alpha}^\JParen  \sigma_{ \alpha }^{(j + 2)}
   \right) .
\end{align}
This Hamiltonian encodes closed boundary conditions.
The numerical simulations (Sec.~\ref{sec_Numerics})
involve periodic boundary conditions.
We extend calculations to periodic boundary conditions 
at the end of the appendix.

%
%
\textbf{Partition function:}
Let us Taylor-approximate the exponential in the NATS:
\begin{align}
   \label{eq_Taylor_Exp}
   e^{ - \beta \left( H^\tot  -  \sum_\alpha  \mu_\alpha  \sigma_\alpha^\tot
                    \right)  }
   = \id  - \beta H^\tot
   + \beta  \sum_\alpha  \mu_\alpha  \sigma_\alpha^\tot
   + O_2 .
\end{align}
The exponential's trace equals $Z$.
The linear terms vanish, as 
$\Tr (  \sigma_\alpha^\JParen  )  =  0$
for all $\alpha$ and $j$. Hence
\begin{align}
   Z   =  2^{\Sites \sites}  +  O_2 .
\end{align}

%
%
\textbf{Inverse temperature:}
$\beta$ follows from the prediction
\begin{align}
   \label{eq_Beta_Predict_App}
   E^\tot
   =  \Tr  \left(  H^\tot
   e^{ - \beta \left( H^\tot  -  \sum_\alpha  \mu_\alpha  \sigma_\alpha^\tot
                    \right) }  \right) / Z .
\end{align}
We substitute in for the exponential from Eq.~\eqref{eq_Taylor_Exp},
then invoke the trace's linearity.
Terms one and three vanish by the Paulis' tracelessness:
\begin{align}
   \label{eq_ETot_Help1}
   E^\tot
   =  - \beta  \Tr  \left( \left[  H^\tot  \right]^2  \right)  /  Z 
   +  O_2 .
\end{align}
Let us evaluate the trace:
\begin{align}
   \Tr  \left( \left[  H^\tot  \right]^2  \right)
   & =  J^2  \sum_{ \alpha,  \alpha' }   \Tr  \Bigg(
   \sum_{j, j' = 1}^{\Sites \sites - 1}
   \sigma_\alpha^\JParen  \sigma_\alpha^{(j+1)}
   \sigma_{\alpha'}^{(j')}  \sigma_{\alpha'}^{(j' + 1)}
   +  \sum_{j = 1}^{\Sites \sites - 1}   
   \sum_{j' = 1}^{\Sites \sites - 2}
   \sigma_\alpha^\JParen
   \sigma_\alpha^{(j + 1)}
   \sigma_{\alpha'}^{(j')}
   \sigma_{\alpha'}^{(j' + 2)}  
   \nonumber \\ & \qquad \qquad \qquad \;
   +  \sum_{j = 1}^{\Sites \sites - 2}
   \sum_{j' = 1}^{\Sites \sites - 1}
   \sigma_\alpha^\JParen
   \sigma_\alpha^{(j + 2)}
   \sigma_{\alpha'}^{(j')}
   \sigma_{\alpha'}^{(j' + 1)}
   +  \sum_{j = 1}^{\Sites \sites - 2}
   \sum_{j' = 1}^{\Sites \sites - 2}
   \sigma_\alpha^\JParen
   \sigma_\alpha^{(j + 2)}
   \sigma_{\alpha'}^{(j')}
   \sigma_{\alpha'}^{(j' + 2)}
   \Bigg) .
\end{align}
Most of the terms vanish, by the Paulis' tracelessness.
In each surviving term, $\alpha = \alpha'$, $j = j'$,
and the second Pauli operator acts on the same qubit as the fourth.
Every Pauli squares to the identity, 
$\left( \sigma_\alpha^\JParen \right)^2  =  \id_2$, so
\begin{align}
   \Tr  \left( \left[  H^\tot  \right]^2  \right)
   & =  J^2  \sum_\alpha 
   \left(  \sum_{j = 1}^{\Sites \sites - 1}  
            +  \sum_{j = 1}^{\Sites \sites - 2}  \right)
   2^{\Sites \sites} \\
   \label{eq_H_Squared}
   & =  3 (2 \Sites \sites - 3)  2^{\Sites \sites}  J^2  .
\end{align}
We substitute into Eq.~\eqref{eq_ETot_Help1}:
\begin{align}
   \label{eq_ETot_Help2}
   E^\tot
   =  - 3 (2 \Sites \sites - 3)  \beta  J^2
   +  O_2 .
\end{align}
Solving for $\beta$ yields Eq.~\eqref{eq_Beta_Predict}.

\textbf{Effective chemical potentials:}
$\mu_\alpha$ follows from the prediction
\begin{align}
   \label{eq_Mu_Predict_App}
   S_\alpha
   =  \Tr  \left(  \sigma_\alpha^\tot  \,
   e^{ - \beta  \left( H^\tot  
                             -  \sum_{\alpha'}  \mu_{\alpha'}  \sigma_{\alpha'}^\tot  \right) 
       }  \right)  /  Z  .
\end{align}
We Taylor-approximate the exponential as in Eq.~\eqref{eq_Taylor_Exp},
then invoke the trace's linearity.
Terms one and two vanish, by the Paulis' tracelessness:
\begin{align}
   \label{eq_Mu_Help1}
   S_\alpha
   =  \left[  \beta  \sum_{\alpha'}  \mu_{\alpha'}
   \Tr \left(  \sigma_{\alpha}^\tot  \sigma_{\alpha'}^\tot  \right)
   +  O_2  \right]  /  Z .
\end{align}
The trace evaluates to
\begin{align}
   \Tr \left(  \sigma_{\alpha}^\tot  \sigma_{\alpha'}^\tot  \right)
   & =  \sum_{j, j' = 1}^{\Sites \sites}
   \Tr  \left(  \sigma_\alpha^\JParen   \sigma_{\alpha'}^{(j')}  \right)
   =  \Sites \sites  \,  2^{\Sites \sites}  \delta_{\alpha \alpha'}  .
\end{align}
In the sum's nonzero terms, the two Pauli operators collide.
We substitute into Eq.~\eqref{eq_Mu_Help1} 
and solve for $\mu_\alpha$:
\begin{align}
   \mu_\alpha
   =  \frac{ S_\alpha }{ \Sites \sites \beta }
   +  O_2 .
\end{align}
Substituting in for $\beta$ from Eq.~\eqref{eq_Beta_Predict}
yields Eq.~\eqref{eq_Mu_Predict}.

\textbf{Small-parameter conditions:}
Inequalities~\eqref{eq_High_T} specify loosely
when our Taylor approximations hold.
More-precise forms for the conditions are presented here.
The conditions follow from calculating second-order corrections,
then demanding that the corrections be much smaller than
the first-order terms:
\begin{align}
   \label{eq_High_T_App}
   \sqrt{3 (2 \Sites \sites - 3)}  \:  
   | \beta | J ,  \;  \;
   \sqrt{\Sites \sites  \sum_\alpha  \mu_\alpha^2}  \:  
   | \beta |  , \;  \;
   \frac{2}{3}  \:
   \frac{ | \beta |  \sum_\alpha  \mu_\alpha^2 }{J}  ,  \;  \;
   6  \:  \frac{\Sites \sites - 2}{ 2 \Sites \sites - 3}  \:
   | \beta |  J ,  \;  \; 
   4  \:  \frac{2 \Sites \sites - 3}{\Sites \sites}  \:
   | \beta |  J
   \ll 1 .
\end{align}

\textbf{Periodic boundary conditions:}
The numerical simulations (Sec.~\ref{sec_Numerics}) 
involve periodic boundary conditions.
The Hamiltonian has the form
\begin{align}
   \label{eq_H_PBCs}
   H^\tot
   = J \sum_\alpha  
   \sum_{j = 1}^{\Sites \sites}  
   \left(   \sigma_\alpha^\JParen  \sigma_\alpha^{(j+1)}
             +  \sigma_\alpha^\JParen  \sigma_\alpha^{(j+2)}
   \right)  .
\end{align}
The site label $j = \Sites \sites + 1$ is defined as $j = 1$,
and $j = \Sites \sites + 2$ is defined as $j = 2$.
Equation~\eqref{eq_H_Squared} changes to 
$\Tr \left( [ H^\tot ]^2 \right)
= 6 \Sites \sites  2^{\Sites \sites}  J^2$,
so $\beta = - \frac{ E^\tot }{ 6 \Sites \sites J^2 }  +  O_2$.
The $\mu_\alpha$ prediction remains unchanged
to first order, though not to second-order.
The small-parameter conditions become
\begin{align}
   \label{eq_Small_Params_PBCs}
   \sqrt{6 \Sites \sites }  \:  | \beta |  J ,  \; \;
   \frac{2}{3}  \:
   \frac{ | \beta |  \sum_\alpha  \mu_\alpha^2 }{J} ,  \; \;
   8 | \beta | J
   \ll 1 .
\end{align}

\section{Quantum state tomography for inferring 
the long-time system-of-interest state}
\label{sec_Tomography}


We aim to observe that $\Sys$, the $\sites$-qubit system of interest,
thermalizes to the NATS.
The following quantum-state-tomography protocol suffices.
A more efficient protocol, that takes advantage of the NATS's form,
might exist.

Let $\vec{\alpha}  =  ( \alpha_1, \alpha_2,  \ldots,  \alpha_\sites )$ 
specify a product 
$\sigma_{\alpha_1}^{(1)}  \otimes
\sigma_{\alpha_2}^{(2)}  \otimes  \ldots  \otimes
\sigma_{\alpha_\sites}^{ (\sites) }$
of Pauli operators.
$3^\sites$ such products exist.
The set of the products' eigenbases
forms a basis for the $\sites$-qubit Hilbert space.
We measure each eigenbasis at the end of
each of $\numTrials$ trials.
Each measurement yields one of $2^\sites$ possible outcomes,
$\ell  =  1, 2, \ldots, 2^\sites$.
If outcome $\ell$ obtains, the projector
$\Pi^{ \vec{\alpha} }_\ell$
projects the state.
Each measurement has a probability
$p_{ \vec{\alpha} } ( \ell | \rho_\Sys )
=  \Tr  ( \Pi^{ \vec{\alpha} }_\ell  \rho_\Sys )$
of yielding outcome $\ell$.
Let $f^{ \vec{\alpha} }_\ell$ denote the frequency with which
measuring $\vec{\alpha}$ yields outcome $\ell$
in our $\numTrials$ trials.
The frequency approximates the probability
with an error $\sim  1 / \sqrt{ \numTrials}$.

From the frequencies, we estimate $\rho_\Sys$.
We can do so by solving the semidefinite program
\begin{align}
   \label{eq_SDP}
   \min_{\rho \, : \,   \rho \geq 0,  \,  \Tr (\rho) = 1}
   \sum_{ \vec{\alpha}  \in  \{ x, y, z \}^\sites }
   \sum_{ \ell = 0 }^{ 2^\sites - 1 }
   \left[ f^{ \vec{\alpha} }_\ell 
           -  \Tr ( \Pi^{ \vec{\alpha} }_\ell  \rho )
   \right]^2  .
\end{align}
Solving this program is equivalent,
in the limit of large $\numTrials$ and so Gaussian noise,
to maximizing the likelihood function
that generated the frequencies.

We can solve the program~\eqref{eq_SDP} efficiently
by recasting the frequencies in terms of expectation values.
Knowing $2^\sites$ probabilities,
we can calculate the expectation values of $2^\sites - 1$
products of Pauli operators and identity operators.
$4^\sites - 1$ such products exist.
They have the form
$\sigma_{m_1}^{(1)}  \otimes
\sigma_{m_2}^{(2)}  \otimes  \ldots  \otimes
\sigma_{m_\sites}^{(\sites)}$.
The $j^\th$ qubit's $m_j = 0, x, y, z$; and 
$\sigma_0^\JParen = \id^\JParen$.
Consider, for example, a system of $\sites = 2$ qubits.
Suppose that we know the four probabilities
$p^{(x, z)}_{ \pm 1,  \pm 1 }$ and $p^{(x, z)}_{ \pm 1,  \mp 1 }$.
We can calculate three expectation values,
$\expval{ \sigma_x  \otimes  \sigma_z }$,
$\expval{ \sigma_x  \otimes  \id }$, and
$\expval{ \id  \otimes  \sigma_z }$.
Hence solving the program~\eqref{eq_SDP} is equivalent to solving
\begin{align}
   \label{eq_SDP2}
   \min_{\rho \, : \,   \rho \geq 0,  \,  \Tr (\rho) = 1}
   \sum_{m  \in  \{ 0, x, y, z \}^\sites}
   \left\{ \expval{  \sigma^{(1)}_{m_1}  \otimes  \ldots  \otimes
                           \sigma^{(\sites)}_{m_\sites}  }
            -  \Tr \left(  \left[  \sigma^{(1)}_{m_1}  \otimes  \ldots  \otimes
                                       \sigma^{(\sites)}_{m_\sites}  \right]  
                             \rho  \right)
   \right\}^2  .
\end{align}
The expectation values 
$\expval{  \sigma^{(1)}_{m_1}  \otimes  \ldots  \otimes
                 \sigma^{(\sites)}_{m_\sites}  }$
are calculated from the measurement data.

The program~\eqref{eq_SDP2} can be solved efficiently as follows~\cite{Smolin_12_Efficient}.
First, we solve the linear inversion problem
\begin{align}
   \min_{\rho}
   \sum_{m  \in  \{ 0, x, y, z \}^\sites}
   \left\{ \expval{  \sigma^{(1)}_{m_1}  \otimes  \ldots  \otimes
                           \sigma^{(\sites)}_{m_\sites}  }
            -  \Tr \left(  \left[  \sigma^{(1)}_{m_1}  \otimes  \ldots  \otimes
                                       \sigma^{(\sites)}_{m_\sites}  \right]  
                             \rho  \right)
   \right\}^2  .
\end{align} 
Then, we impose the positive-semidefinite
and trace constraints.

\section{Protocol's robustness with respect to experimental error}
\label{sec_Robust}

In~\cite{Fukuhara_13_Microscopic}, a nearly isotropic Heisenberg model
is effected with a Bose-Hubbard Hamiltonian
in the hardcore limit.
The Hamiltonian has the form
\begin{align}
   \label{eq_H_BH}
   H_\BH = - J_{\rm ex}  \sum_j  \left[
   2 \left(  \sigma_{+z}^\JParen  \sigma_{-z}^{(j+1)}
               +  \sigma_{-z}^\JParen  \sigma_{+z}^{(j+1)}   \right)  
   +  \Delta  \sigma_z^\JParen  \sigma_z^{(j+1)}   \right] .
\end{align}
Again, we have ignored factors of $\hbar / 2$. 
$J_{\rm ex}$ denotes the energy scale,
and $\Delta$ denotes the isotropy parameter.
$H_\BH$ becomes an isotropic Heisenberg model when $\Delta = 1$.
When $\Delta \neq 1$, angular momenta associated with different axes
hop at different rates.
$H_\BH$ consequently conserves only $\sigma_z^\tot$,
not $\sigma_x^\tot$ and $\sigma_y^\tot$.
An isotropy parameter of $\Delta = 0.986$ was achieved in the experiment.

We investigated our protocol's robustness with respect to this error.
We simulated evolution under a Hamiltonian 
that resembles~\eqref{eq_H_BH}
but that encodes next-nearest-neighbor couplings:
\begin{align}
   \label{eq_H_BH_2}
   \tilde{H}_\BH
   & =  -  J_{\rm ex}  \left[  \sum_{j = 1}^{\Sites \sites}  
   \left(  \sigma_x^\JParen  \sigma_x^{(j+1)}
            +  \sigma_y^\JParen  \sigma_y^{(j+1)}
            +  \Delta  \sigma_z^\JParen  \sigma_z^{(j+1)}  \right)
   +  \sum_{j = 1}^{\Sites \sites} 
   \left(  \sigma_x^\JParen  \sigma_x^{(j+2)}
            +  \sigma_y^\JParen  \sigma_y^{(j+2)}
            +  \Delta  \sigma_z^\JParen  \sigma_z^{(j+2)}  \right)
   \right] .
\end{align}
As in Sec.~\ref{sec_Numerics}, we simulated periodic boundary conditions.
We chose for the nearest-neighbor and next-nearest-neighbor terms
to have the same $\Delta$.
We focused on a 1\% anisotropy and set $J_{\rm ex} = 1$.
To mitigate the error, we implemented the scheme in~\cite{Viola_99_Universal}
(Sec.~\ref{sec_Proposal}):
The evolution time $t = 2^{\Sites \sites}$ was split into steps of duration
$dt = t / (3 \times 2^{\Sites \sites} + 1)$.
After each time step, the system underwent a 90$^\circ$ rotation.
(Qubits can be rotated experimentally with microwave pulses.)
The $x$-axis was rotated into the $y$-axis,
then into the old $z$-axis,
and then returned to its original orientation.
This cycle was then repeated.

\begin{figure}[hbt]
\centering
\includegraphics[width=.5\textwidth, clip=true]{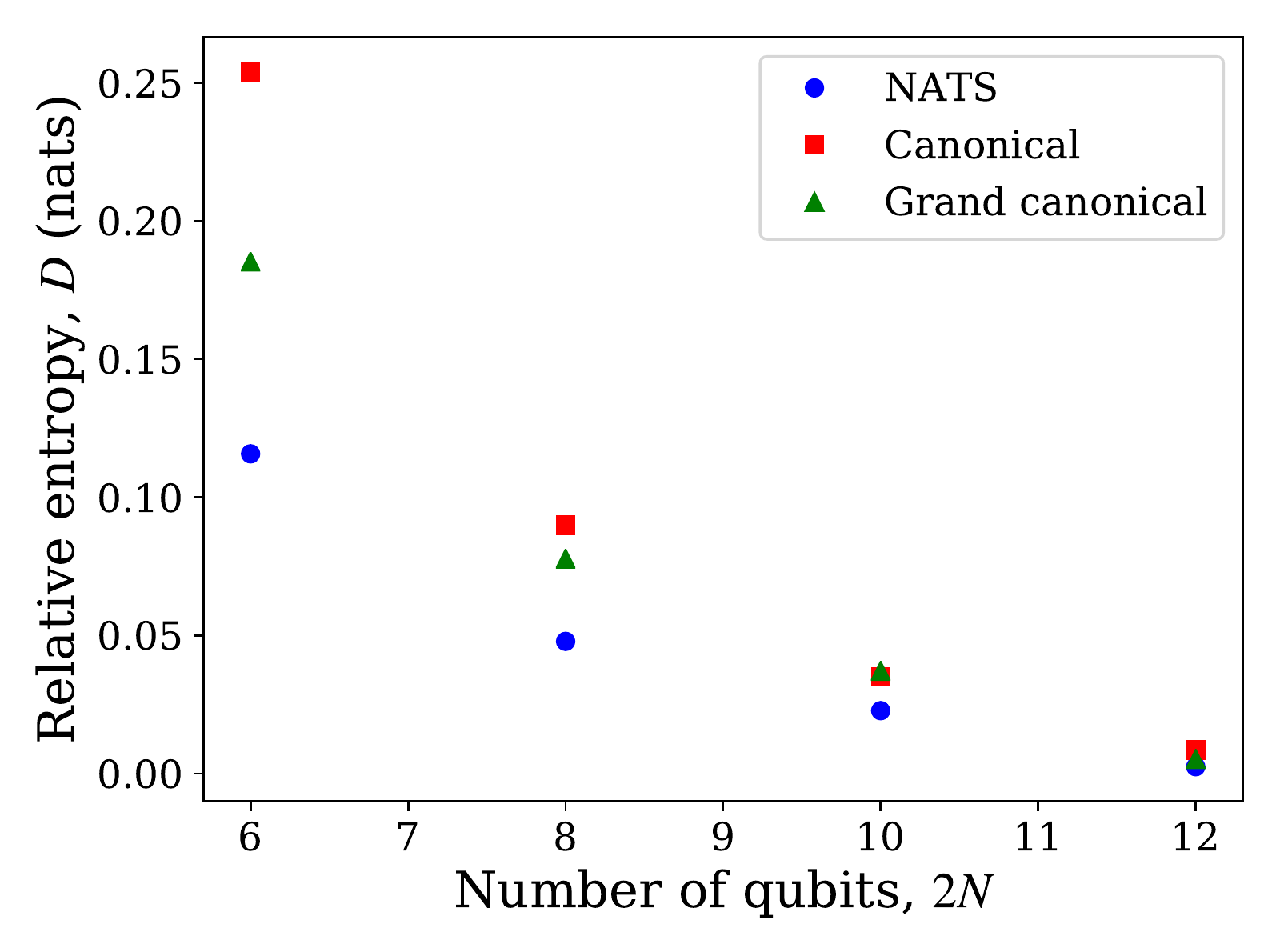}
\caption{\caphead{Protocol's robustness with respect to anisotropy:} 
Experimental implementations of the Heisenberg Hamiltonian~\eqref{eq_HTot}
may involve anisotropic couplings.
Evolution under the Hamiltonian~\eqref{eq_H_BH_2} was simulated with
an isotropy parameter of $\Delta = 0.99$,
in mimicry of the experiment in~\cite{Fukuhara_13_Microscopic}.}
\label{fig_Robust}
\end{figure}

Figure~\ref{fig_Robust} shows the resulting relative entropies.
Each state was calculated from $H^\tot$, as though the error were absent.
For example, $\rho_\NATS$ continues to have the form in Eq.~\eqref{eq_NATS}.
The NATS prediction remains the most accurate, 
despite the simulated experimental error.

%
%
\section{NATS adaptation of Deutsch's argument for studying the ETH}
\label{app_Deutsch}


Deutsch's original ETH paper~\cite{Deutsch_91_Quantum}
offers another lens through which to view our NATS protocol.
The ETH describes a closed quantum many-body system's 
thermalization to a canonical state.
Quantum systems were known to thermalize to the canonical state
by exchanging heat with external baths.
Did the ETH not therefore recapitulate well-known physics?
No, Deutsch argued: Different mechanisms drive
the two thermalization processes.
Similarly, consider placing a spin system $\Sys$ in a magnetic field
$\vec{B} = \sum_{\alpha}   \mu_\alpha   \hat{\alpha}$
and in contact with an inverse-temperature-$\beta$ bath.
$\Sys$ thermalizes to a state identical to the NATS,
$\rho_\triv :=
e^{ - \beta ( H^\Sys - \sum_\alpha \mu_\alpha \sigma_\alpha^\Sys ) }
/ Z_\NATS^\Sys$.
This thermalization is well-understood.
Yet the NATS remains nontrivial:
Different physics drives the two thermalizations, as in Deutsch's argument.
A classical external field thermalizes spins to $\rho_\triv$.
Exchanges of noncommuting charges 
within a closed, isolated quantum system
thermalizes spins to the NATS.
As ETH thermalization merits study,
so does NATS thermalization.
NATS thermalization arguably demands more,
highlighting nonclassical noncommutation.

\end{appendices}

%
%
\bibliographystyle{h-physrev}
\bibliography{Noncommq_refs}

\end{document}